\documentclass[journal]{IEEEtran}
\IEEEoverridecommandlockouts
\usepackage{cite}
\usepackage{amsmath,amssymb,amsfonts}
\usepackage[dvipdfmx]{graphicx}
\usepackage{textcomp}
\usepackage{xcolor}
\usepackage{algpseudocode}
\usepackage{caption}
\usepackage{subcaption}

\usepackage[ruled,vlined, linesnumbered]{algorithm2e}
\def\BibTeX{{\rm B\kern-.05em{\sc i\kern-.025em b}\kern-.08em
    T\kern-.1667em\lower.7ex\hbox{E}\kern-.125emX}}

\newcommand{\ADP}{\ensuremath{\text{ADP}}}
\newcommand{\ADPPARAM}{\ensuremath{\psi_{\ADP, m}}}
\newcommand{\ADPM}{\ensuremath{\Pr\left(\text{SINR}_{b,m}\geq\tau  ~|~ \ADPPARAM\right)}}
\newcommand{\JDP}{\ensuremath{\text{JDP}}}
\newcommand{\JDPPARAM}{\ensuremath{\psi_{\JDP, m}}}
\newcommand{\JDPM}{\ensuremath{\Pr\left(\text{SINR}_{b,m}\geq\tau,\text{SINR}_{v,m}\geq\tau~|~d_{b,v},\JDPPARAM\right)}}
\usepackage{soul}
 
\begin{document}

\newtheorem{proposition}{Proposition}

%\title{{Base  Station Placement and Frequency Assignment in Multiband Ultra-Narrowband Systems for Massive IoT Access}\\ 
\title{{Multiband Massive IoT: A Learning Approach to Infrastructure Deployment}\\
% Optimal Band Assignment for UNB LPWA Networks with Multiband Access\\
\thanks{Enes Krijestorac and Danijela Cabric are with the Electrical and Computer Engineering Department, University of California, Los Angeles, CA 90095, USA. e-mail: enesk@g.ucla.edu, danijela@ee.ucla.edu.}
\thanks{Ghaith Hattab was with the Department of Electrical and Computer
Engineering, University of California at Los Angeles, Los Angeles,
CA 90095-1594 USA. He is now with Apple Inc., Cupertino, CA 95014 USA
(e-mail: ghattab@ucla.edu).}
\thanks{Petar Popovski is with the Department of Electronic Systems, Aalborg
University, 9220 Aalborg, Denmark (e-mail: petarp@es.aau.dk).}
\thanks{The work of Petar Popovski has been in part supported by the Danish Council for Independent Research, Grant Nr. 8022-00284B SEMIOTIC, and the Villum Investigator Grant “WATER” from the Velux Foundation, Denmark.}
}

\author{\IEEEauthorblockN{Enes~Krijestorac, Zheang Huai, Ghaith Hattab, Petar Popovski, Danijela Cabric}
\IEEEauthorblockA{{Electrical and Computer Engineering Department,} {University of California, Los Angeles} \\ 
Los Angeles, USA\\
		enesk@ucla.edu,
		samerhanna@ucla.edu, danijela@ee.ucla.edu 
}
}
\author{
    \IEEEauthorblockN{Enes~Krijestorac,~\IEEEmembership{Student~Member,~IEEE}, Ghaith Hattab,~\IEEEmembership{Member,~IEEE},~Petar Popovski,~\IEEEmembership{Fellow,~IEEE},~and~Danijela Cabric,~\IEEEmembership{Fellow,~IEEE}}
    
}

\maketitle

\IEEEpubid{\begin{minipage}{\textwidth}\ \\[30pt] \centering
  \copyright 2022 IEEE. Personal use of this material is permitted. Permission
from IEEE must be obtained \\
for all other uses, in any current or future
media, including reprinting/republishing this material for advertising or \\
promotional purposes, creating new collective works, for resale or
redistribution to servers or lists, \\or reuse of any copyrighted
component of this work in other works.
\end{minipage}}

\begin{abstract}
%Low-power  wide-area  (LPWA) networking is a recent paradigm in wireless  networking, intended to support access from a massive number of Internet of Things (IoT) devices. 
%Ultra-narrowband (UNB) LPWA solutions apply the ultra-narrowband transmissions, which enable demodulation at very low received power.
%Low-power  wide-area  (LPWA) networks is a recent paradigm in wireless  networking, intended to support access from a massive number of Internet of Things (IoT) devices. A variant is  the  ultra-narrowband (UNB) network, where the transmission bandwidth is often smaller than the uncertainty of the carrier frequency. 
We consider a novel  ultra-narrowband (UNB) low-power wide-area  network (LPWAN) architecture design for uplink transmission of a massive number of Internet of Things (IoT) devices over multiple multiplexing bands. An IoT device can randomly choose any of the multiplexing bands to transmit its packet. Due to hardware constraints, a base station (BS) is able to listen to only one multiplexing band. %That hardware constraint is mainly due to the complexity of performing fast Fourier transform (FFT) at a very small sampling interval over the multiplexing bands in order to counter the uncertainty of IoT device frequency and synchronize onto transmissions.
Our main objective is to maximize the packet decoding probability (PDP) by optimizing 
the placement of the BSs and frequency assignment of BSs to multiplexing bands.
%We develop two training-based approaches that require estimation of a finite number of parameters for optimal assignment and placement of BSs. 
We develop two online approaches that adapt to the environment based on the statistics of (un)successful packets at the BSs.
% EDITS IN RESPONSE TO PROF CABRIC
%The  approach is model-based and has a low training complexity. The alternative approach has is based on a looser model of the IoT network and the environment at the cost of a higher training complexity. We discuss when either of the approaches is appropriate. 
%The two approaches differ in the strictness on the assumed models of the environment. 
The first approach is based on a predefined model of the environment, while the second approach is measurement-based model-free approach, which is applicable to any environment. The benefit of the model-based approach is a lower training complexity, at the risk of a poor fit in a model-incompatible environment. The simulation results show that our proposed approaches to band assignment and BS placement offer significant improvement in PDP over baseline random approaches and perform closely to the theoretical upper bound.
\end{abstract}


\section{Introduction}
\IEEEPARstart{T}{he} Internet of Things (IoT) has the potential to change the technological landscape and bring great economical and societal benefits. The success of IoT on a large scale depends on several key enabling technologies, one of which is wireless communication.
Low-power wide-area (LPWA) networks are a new paradigm of wireless networking that is expected to become one of the key drivers of massive IoT \cite{raza2017low}. 
Compared to the legacy technologies, such as cellular and short-range wireless networks, {the benefits offered by LPWA networks include} wide-area connectivity for low-power and low-data-rate devices and low capital expenditure due to the use of the unlicensed spectrum. 
To enable long-range connectivity, LPWA networks primarily use sub-1GHz bands due to the favorable propagation conditions. Ultra-narrowband (UNB) LPWA solutions apply the ultra-narrowband transmissions, which enable demodulation at a very low received power. Furthermore, UNB LPWA networks normally rely on simple ALOHA-like access protocols, where IoT devices avoid associating and synchronizing with any UNB {base station} (BS); in essence, IoT devices operate in a broadcast mode, transmitting their packets at arbitrary time and frequency. Normally, the BSs operate in a decentralized manner and packet decoding occurs at the BSs, as opposed to at a central server where the received symbols from all BSs in the network could be combined. Therefore, the packet is successfully transmitted if any BS decodes any of the packet transmissions.

{Due to the extremely low bandwidth (hundreds of Hz), the frequency drift of the local oscillator in the commodity hardware becomes comparable to the transmission bandwidth, thus rendering slotted access infeasible.} UNB networks solve this problem by allowing the devices to transmit in an unslotted manner, while the receiving BSs do the task of accurately syncing on to a signal in frequency. 
This is accomplished by performing fast Fourier transform (FFT) at a very fine sampling interval over the entire bandwidth of the multiplexing band, which is the portion of the spectrum across which UNB transmissions occur. 
Naturally, a UNB network would benefit from a wider multiplexing band. 
%However, since the complexity of the FFT scales as $O(B\log B)$ with bandwidth $B$ of the multiplexing band, $B$ has a feasibility limit.
However, since the complexity of the FFT scales with the bandwidth of the multiplexing band, the multiplexing band has a feasibility limit on its bandwidth.
One way to introduce more frequency diversity would be to use several multiplexing bands with each BS associated to one band and IoT devices transmitting freely across any band. 
This would keep the capital expenditure the same since the BSs would still use the same hardware, only tuned to different multiplexing bands, while the capacity of the network could potentially increase. Indeed, in \cite{hattab2018spectrum, hattab2020spectrum} it has been shown that the capacity does increase by applying this paradigm at no additional cost. However, the scope of \cite{hattab2018spectrum, hattab2020spectrum} does not consider optimization of the BS infrastructure, leaving room for further enhancements.

\IEEEpubidadjcol

Optimizing the {deployment of the} BS infrastructure, namely the BS placement and frequency band assignment, is precisely the focus of this paper.
%The main objective of the said optimization is to maximize the probability of decoding of a packet (PDP) irrespective of the source. 
The BS placement and channel assignment are jointly optimized, however, the approaches that we will propose naturally extend to optimizing channel assignment without placement and vice versa. {We have developed an approach for joint optimization of BS placement and band assignment, however, this approach can be applied to solve the problem of
band assignment only, when the placement is predetermined. It can also be applied
to solving for placement only, when there is a single band.}
Moreover, we assume that some BSs may be present in the environment already and the network is expanded by introducing additional BSs. In that case, the optimization  will include channel assignment for the current BSs as well as both channel assignment and placement for the new BSs.  
{The objective is to maximize the probability of decoding of a packet (PDP), irrespective of the source {IoT device}}.

We cast the problem as an integer non-linear problem (INLP), for which we then find an approximate solution. In order to implement this solution, we are faced with learning the \emph{average decoding probability (ADP)} at several locations in the network and the \emph{joint decoding probability (JDP)} between several pairs of locations in the IoT network.
{JDP between a pair of locations is defined as the probability of successfully decoding a transmission from an IoT device at both locations in the pair.}
%\pp{PP: While ADP is intuitively clear, please put a sentence to explain the rationale behind JDP.} In order to learn ADPs and JDPs, we propose two alternative approaches: a model-based and a measurement-based approach. 
%The model-based approach requires fewer measurements, however it is based on stochastic models which the environment should match. 
% Moved this part into contributions list
The model-based approach requires fewer measurements, however, it is based on stochastic models of device placement, channel, and interference, which the environment should match. 
%The measurement-based approach is based on very loose assumptions at the cost of a more extensive training phase, and it is more suitable for channel assignment than BS placement. Our main contributions can then be summarized to these key points:
{The measurement-based approach directly estimates the required JDPs and ADPs.} It involves no assumptions on device placement, channel, and interference at the cost of a more demanding training phase. Due to its training complexity, this approach is more suitable for optimizing only channel assignment when the placement { is not optimized, due to, for example, infrastructure constraints that decide the BS placement}. 
In both approaches, we do not assume to know the locations of IoT devices, rather, our optimization is based on the information collected by the {BSs that are receiving the packets.}

{Our main contributions can be summarized as follows:}
\begin{itemize}
    \item We formulate joint BS placement and frequency assignment for maximizing PDP as an optimization problem and derive an approximate solution, which we show to require the estimation of ADP and JDP at several locations in the network. 
    \item We propose a model-based approach for predicting ADP and the JDP at locations for which we do not have any measurements. First, we use stochastic geometry to derive models of ADP and JDP as a function of the location. 
    %We then use the derived models to develop an algorithm for prediction of ADP and JDP at unvisited locations.
    We then use the derived models to develop an algorithm for prediction of ADP and JDP for any location in the network.
    %\item We develop an algorithm for efficient direct learning of ADP and JDP at required locations with loose assumptions on the characteristics of the IoT network and the environment. 
    % I added the explanation in the paragraph above
    \item We develop an algorithm for efficient direct learning of ADP and JDP for candidate locations for the BSs, which is the basis for the approach that we refer to as the measurement-based approach.  
\end{itemize}

%Also, in our approach we do not assume to know the locations of IoT devices, rather, our optimization is solely based on the information collected by the receivers, i.e. the BSs.

%The remainder of the paper is organized as follows. In Sec. \ref{sec:related-work} we summarize the related work and highlight the importance of this work. 
% Moved to an earlier paragraph
\emph{Organization:}
The remainder of the paper is organized as follows. In Sec. \ref{sec:related-work}, we summarize the related work and state the novelty compared to related work. 
In Sec. \ref{sec:iot-network-model}, we describe the model of the UNB IoT network and the incumbent devices. In Sec. \ref{sec:problem}, we mathematically describe the problem of BS placement and BS band assignment and pose it as an INLP. We then show how the problem can be relaxed such that it can be solved by knowing ADPs and JDPs at a number of locations in the environment. 
%In Sec. \ref{sec:model-based-approach}, we describe the model-based approach for solving the problem posed in Sec. \ref{sec:problem}. In Sec. \ref{sec:training}, we describe the training procedure for the approach described in Sec. \ref{sec:model-based-approach}. In Sec. \ref{sec:meas-approach}, we describe the measurement-based approach for solving the problem posed in Sec. \ref{sec:problem}. In Sec. \ref{sec:sim-results} we use simulations to verify and compare our approaches and in Sec. \ref{sec:conclusions} we provide conclusions and directions for future work.
In Sec. \ref{sec:model-based-approach}, we describe the model-based approach and in Sec. \ref{sec:training}, we describe its training procedure.  In Sec. \ref{sec:meas-approach}, we describe the measurement-based approach. In Sec. \ref{sec:sim-results}, we run simulations to evaluate and compare our approaches, and in Sec. \ref{sec:conclusions}, we provide conclusions and directions for future work.

{
\emph{Notations:} Scalars, vectors, and matrices are denoted by regular, bold lower-case, and bold upper-case letters, respectively. The $(i, j)$-th element of $\mathbf{A}$ is denoted by $[\mathbf{A}]_{i, j}$. A two-element tuple is denoted as $(a,b)$. An uniform random distribution in a closed interval $[a,b]$ is denoted as $\mathcal{U}[a,b]$. The set of all vectors of size $N$ with binary valued entries is denoted as $\mathbb{Z}_2^N$. The function $\mathbf{1}_A$ is an indicator function equal to 1, if the logic statement $A$ is true, and 0 otherwise. $|\mathcal{A}|$ is the Lebesque measure of a continuous set $\mathcal{A}$ or the cardinality of a discrete set $\mathcal{A}$. $\text{vec}(\mathbf{A})$ flattens a matrix into a column vector.
}

\section{Related Work}
{\subsection{Our previous work}}
{The theoretical benefits of multiband UNB LPWAN networks have been analyzed in \cite{hattab2018spectrum, hattab2020spectrum} using a stochastic geometry framework and simulation studies.
Using multiple bands reduces IoT collisions and interference with incumbents. However, to fully exploit these gains when each BS is restricted to listen to one of the bands, it is necessary to optimize the BS-band selection policy.
In \cite{krijestorac2020band}, we proposed an algorithm for optimal BS-to-band assignment in a LPWA UNB multi-band network. This paper is an extension of that work with the focus expanding to joint BS placement and band assignment optimization. 
%As we will show in this paper, the extensions to BS placement is an important direction in infrastructure optimization in a multi-band UNB LPWAN and needs to be optimized jointly with band assignment. 
One of the two main proposed algorithms that we will discuss in this paper, the measurement-based approach, is based on our work in \cite{krijestorac2020band}, but in this work it has been extended to be applicable for placement optimization. The second approach that will introduce in this paper, the model-based approach is entirely novel.
}
{\subsection{Existing prior literature}}
{While the channel assignment and BS placement problems were previously researched in the context of various wireless technologies such as LPWANs, Wi-Fi networks \cite{ling2006joint}, cellular networks \cite{uygungelen2011graph, narayanan2001static, chae2011radio}, and cognitive unlicensed radio networks \cite{ahmed2014channel}, to the best of our knowledge, no problem setting has considered the problem of optimal BS placement and band assignment in a multi-band UNB LPWAN. 
% This is a redundant sentence
%\red{\st{Therefore, the solutions developed for channel assignment and BS placement for the said wireless technologies, would not apply to the problem at hand.}} 
The distinctions arise due to two characteristics:  (i) We consider a multi-band network in which IoT devices are able to transmit in any band but the receivers (BSs) are tuned to only one band {over a long period of time}; and (ii) we assume that there is no association between devices and BSs, i.e. any BS that receives a transmission is expected to decode it, which is the case for a limited number of network architectures. While LPWANs such as LoRa often satisfy characteristic (ii), the characteristic (i) is not met. Other technologies, such as Wi-Fi or cellular, in general neither meet (i) nor (ii). 
%\pp{PP: Another key aspect of this work is \emph{long-term} optimization of the network, as it evolves over a longer history, rather than the usual optimization in wireless networks that happens for a very limited time interval, often a single instant in time (e.g. in scheduling).} \red{EK: That is true, but criteria (i) and (ii) would already exclude networks that involve short-term scheduling.}
}
{
In the context of LPWANs, there are several works that consider BS placement optimization but not channel assignment. In \cite{ousat2019lora}, the authors consider the optimal placement of BSs for LoRa networks, however, their main objective is to minimize the capital-expenditure costs and maximize energy efficiency of the devices. Furthermore, it is assumed that the locations of the IoT devices are known which makes it possible to fairly reliably predict the PDP of a device given a particular BS placement, which is taken advantage of in their approach. We do not make the same assumptions since it is unlikely that all devices served by the BSs will be equipped with a GPS or other localization technologies due them often being energy limited.
%In \cite{matni2019optimal}, a similar objective is considered for the general category of LPWAN networks. \cite{tian2018optimized} considers BS placement optimization with the objective of maximizing the packet delivery ratio, equivalent to PDP in this paper, however, it assumed that the locations of the IoT devices are known, which is not one of the assumptions that we make since estimating the locations of IoT devices is challenging and not all IoT devices that are deployed are equipped with a GPS module. 
In \cite{matni2019optimal}, a similar objective is considered for the general category of LPWAN networks with known IoT device locations. The work in \cite{tian2018optimized} considers placement when gateways perform interference cancellation with the objective of maximizing the packet delivery ratio, equivalent to PDP in this paper, however, it is assumed that the locations of the IoT devices are known. In summary, no  approaches that were previously developed for placement in LPWANs would apply to the problem that we consider. Furthermore, to the best of our knowledge, no works in LPWANs consider channel assignment on the BS side because the BSs are normally assumed to operate on a single channel.

\label{sec:related-work}
{\subsection{Overview of current technologies}}
{Several commercial UNB LPWAN {technologies} are available on the market at the time of this writing, such as Sigfox \cite{sigfox_usa_2020}, WAWIoT NB-Fi \cite{wawiot},  and Telensa \cite{telensa}. For a more detailed survey of various LPWAN technologies, including UNB, the reader is referred to \cite{finnegan2018comparative}. Here, we briefly summarize the main characteristics cited in the technical specifications of these technologies that will be relevant to this thesis. First, the expected range of reliable communication is up to 10 miles. The uplink packet size payload is up to 12 bytes, depending on the type of data being transmitted, and the bandwidth of packet transmission ranges from 50Hz to 600Hz across technologies. Multiplexing band bandwidth ranges from 200kHz to 500kHz and the bands are located in the sub 1GHz unlicensed ISM bands for all existing commercial solutions.}

%For example, channel assignment in 802.11 WLANs was previously investigated (see\cite{ling2006joint} and references therein). 
%However, in WLANs, the user equipment (UE) is associated with a single access point while for the use case of UNB networks, the transmissions are broadcasted to all BSs. Channel assignment problems also appear in cognitive radio networks \cite{ahmed2014channel}, however, in none of the works that we surveyed, there exists a similar infrastructure to that of UNB LPWA networks. 
%Finally, frequency reuse has received a great deal of attention with respect to cellular networks \cite{uygungelen2011graph, narayanan2001static, chae2011radio}, but in these problems, the goal is usually to assign frequency bands such that BSs that are in the neighborhood of one another do not receive the same band. This approach would not apply to our problem as its primary goal is to maximize the frequency utilization, while we are mainly concerned with maximizing packet decoding probability (PDP). 

%One exception is our previous work \cite{krijestorac2020band}, where we proposed an algorithm for optimal BS channel assignment in a LPWA UNB multi-band network. This paper is an extension of that work with the focus expanding to joint BS placement and channel assignment optimization.

\begin{figure}[t]
    \centering
    \includegraphics[width=.9\linewidth]{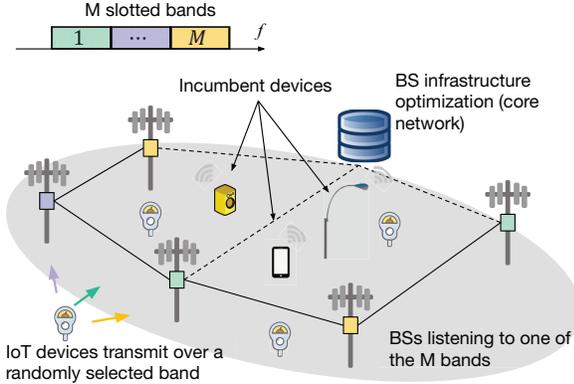}
    %\vspace{-5pt}
    \caption{The system model of the LPWAN IoT network.}
    \label{fig:my_label}
    %\vspace{-10pt}
\end{figure}

\section{Model of the IoT network}
\label{sec:network-model}
% TODO: Relate this system model to some real examples of IoT technologies.
\label{sec:iot-network-model}
In this section, we describe the IoT network considered for the BS placement optimization and BS frequency assignment. 
%We will focus only on a high level description of the network, which will give us enough foundation to state the target problem and propose our solutions.
%As explained in the introduction, we are interested in the planning of the operation of a LPWA UNB IoT network operating across multiple unlicensed bands below $1$ GHz.
{We are interested in the planning of the operation of a LPWA UNB IoT network operating across multiple sub-1GHz unlicensed bands.}
%Unlike most other works that analyze UNB LPWANs, where the interference is assumed to come only from other devices of the same technology, the IoT network is assumed to share the unlicensed band with other incumbent technologies such as LoRa, Zigbee or Wi-Fi. We assume that the incumbent technology can be modeled with some regular characteristics. 
% https://en.wikipedia.org/wiki/IEEE_802.11ah#Chipset

The interference is assumed to come not only from other devices of the same technology, as the IoT network is assumed to share the unlicensed band with other incumbent technologies such as LoRa, Zigbee or 802.11 WLANs. 
% Details of the model described later
%We will approximately model the incumbent devices to have a uniform placement distribution and Poisson packet arrival times. However, different incumbent devices may occupy different multiplexing bands and therefore their 

{
Among the assumptions made in this section, only the assumptions described in Sec.  \ref{sec:iot-network-model-a}
are used by the measurement-based approach while others are only necessary for model-based approach.
} 
%\emph{Medium access mechanisms:} 
\subsection{UNB network topology and transmission model}
\label{sec:iot-network-model-a}
We denote the number of BSs present in the network as $B$.  There are $M$ multiplexing bands, each with bandwidth $W$. {For ease of exposition, we assume a constant $W$ across bands $m\in\left\{1,..,M\right\}$, even though our analysis is not dependent on this assumption.} IoT devices use ALOHA random access protocol to access the bands and transmit signals at power $P$, occupying a bandwidth $w$ with a random center frequency. 
%  For the temporal generation of IoT traffic, UNB devices randomly transmit packets over time at a rate of $N$ packets per hour.
Each packet is repeated $R$ times, consecutively over time, yet randomly hopping from one frequency to another {across any of the $M$ bands}, as shown in Fig. \ref{fig:my_label}. The packet transmission duration is $T$. We label the transmissions of IoT devices with a tuple $(n,r)$, where {$n\in\left\{1,2,..,\infty\right\}$} is the packet index and $r\in\left\{1,..,R\right\}$ corresponds to the repetition number of the packet. Therefore, the probality of a device being active is $\frac{NRT}{1~h}$.
{In UNB LPWANs such as Sigfox, IoT devices are allowed to transmit at an average rate of up to $N$ packets per hour \cite{sigfox_usa_2020}, although they may transmit less in order to conserve energy. We assume that all devices are transmitting $N$ packets per hour and that packets are transmitted as soon as they arrive into the queue, since there is no listen-before-talk mechanism. Hence, we model packet arrivals as a Poission process, with independent arrivals across devices.}   

We denote the time of a transmission $(n,r)$ as $t(n,r)$. 
The transmission signals are extremely narrowband and the channel access is assumed to be unslotted, therefore we model the carrier frequency $f(n, r)$ as a uniform random distribution $\mathcal{U}\left[\frac{w}{2},WM-\frac{w}{2}\right]$. 
Similarly, since the devices randomly transmit across any of the multiplexing bands, we can model the band selected for a particular packet transmission $\beta(n,r)$ as a discrete uniform random variable that takes on the values $m  \in \{1,...,M \}$. The probability distributions across each band is approximately $\Pr(\beta(n,r)=m)=\frac{1}{M}$, since $w \ll W$.

\subsection{Interfering networks}
We will assume a similar regular access pattern for incumbent devices with $N_I$ packets per hour, $R_I$ repetitions, power $P_I$, bandwidth $w_I$ and transmission time $T_I$ . {The transmit probability of incumbent devices is then $\frac{N_I T_I}{1~h}$. We also assume that transmissions can be modeled by a Poisson point process that operates independently across devices. } {This assumptions is true only when the queues at each device are backlogged or packets arrivals at each device are sporadic.} {Otherwise, the analysis would be very difficult due to spatio-temporal correlation between incumbent transmissions \cite{zhong2020spatio}.} 
%\pp{PP: I would suggest a different notation for the incumbent: $N_I$ packets per hour, $R_I$ repetitions, power $P_I$, bandwidth $w_I$ and transmission time $T_I$}

%We choose this general model since it allows us to make problem analysis more tractable, even though it may not capture all of the properties of possible incumbent technologies, e.g.\red{,} backoff periods and retransmission in 802.11 protocols.  

%We will assume that the incumbent devices are confined to transmit within a single multiplexing band $m$. 
%Without the loss of generality, we will assume that the incumbent devices are confined to transmit within a single multiplexing band $m$. If a particular device is occupying several bands, it can be modeled by separate virtual devices. For simplicity, however, we assume that the interference across multiplexing bands is uncorrelated.
{To simplify our analysis, we further assume that the interference across multiplexing bands is uncorrelated. With this assumption, if a particular device is occupying several bands, it can only be approximately modeled by separate uncorrelated virtual devices.}

The number of incumbent devices operating on each multiplexing band $m$ may not be the same, therefore IoT devices on some bands may experience a higher incumbent interference than on others. 
%The assumption of unequal incumbent interference across different frequencies is consistent with measurement studies performed in IoT networks. \cite{vejlgaard2017interference}.
The assumption of unequal incumbent interference across different frequencies is consistent with measurement studies performed in IoT networks \cite{vejlgaard2017interference}, which show non-uniform interference power distribution in sub-1 GHz ISM bands.
%\emph{Network topologies:}

%We consider a randomly distributed spatial topology of BSs, IoT devices, and incumbent devices distributed across some area $\mathcal{A}$. 
%We consider a uniform distribution of BSs, IoT devices, and incumbent devices distributed across some area $\mathcal{A}$. 
%There are $D$ UNB IoT devices that are randomly distributed across $\mathcal{A}$ and their location is assumed to be unknown, but we assume that their locations are sampled from a uniform distribution. 

\subsection{UNB and Interfering devices' topology}
There are $D$ UNB IoT devices {across an area of interest} $\mathcal{A}$ whose locations are assumed to be unknown, however, we assume that their locations are sampled from a uniform distribution. 
{The area $\mathcal{A}$ contains all the IoT devices for which we aim to maximize the PDP by optimizing the BS deployment.}
Modeling the distribution of the locations of wireless devices as uniform is justified in many instances, including IoT networks. The number of incumbent devices on each multiplexing band $m$ is $D_{I,m}$ and their locations are also sampled from a uniform distribution across $\mathcal{A}$. 

\section{Problem statement and proposed solutions}
\label{sec:problem}
In this section, we define the problem of optimal BS placement and the optimal assignment of BSs to multiplexing bands. 

\subsection{Problem statement}
The main objective of the algorithm will be to maximize the probability of decoding a packet (PDP) for a typical device in the network. 
There are $B \geq 0$ currently deployed BSs and we assume that there is a need to expand the BS infrastructure by installing new BSs in $\mathcal{A}$. 
The network will be expanded by introducing additional $\Delta B \geq 0$ BSs to the network, and we need to find an optimal placement for these BSs and their multiplexing band assignment. When new BSs are installed, the band assignment of the currently deployed BSs may need to be updated. We assume $\Delta B$ is pre-determined by the resource availability of the network operator { and not  optimized}.  When $\Delta B=0$, then the problem reduces to the optimal assignment of BSs to bands. When installing new BSs, we assume that there are $C \geq \Delta B$ candidate locations for new BSs.  This would be determined by practical constraints such as connectivity to the core network or real-estate considerations. Among the $C$ candidate locations, $\Delta B$ locations will be populated by new BSs.

Let us introduce an extended set of BS locations that includes the locations currently occupied by BSs and the candidate locations for the new BSs $b \in \{ 1,\dots,B,\dots B+C\}$, where the locations $b \in \{ 1, \dots, B\}$ correspond to the locations currently occupied by the BSs.

We will be solving for the binary assignment variable $\mathbf{X}\in \mathbb{Z}_2^{(B+C) \times M}$, where
\begin{equation}
    [\mathbf{X}]_{b,m}=\begin{cases}
    1 & \text{A BS is placed at }b\text{ and assigned to band }m\\
    0 & \text{No BS at location }b\text{ and band }m
    \end{cases}.
\end{equation}

% Our objective is to maximize the expected PDP defined in the Eq. \ref{eq:main_obj} of a typical packet $n$, given an assignment $\mathbf{X}$. 

% \begin{equation}
%     P_{\text{PDP}}(\mathbf{X})=Pr\left(\bigcup_{r}\bigcup_{b}\left[\mathbf{X}\right]_{b,\beta(n,r)}\text{SINR}_{b,\beta(n,r)}(n,r)>\tau\right)
%     \label{eq:main_obj}
% \end{equation}
% Rephrased this slightly to make it more clear

Given an assignment $\mathbf{X}$, our objective is to maximize the average PDP of a packet $n$, defined as follows: 
\begin{equation}
    P_{\text{PDP}}(\mathbf{X})=\Pr\left(\bigcup_{r}\bigcup_{b}\left[\mathbf{X}\right]_{b,\beta(n,r)}\text{SINR}_{b,\beta(n,r)}(n,r)>\tau\right)
    \label{eq:main_obj}
\end{equation}

%\pp{PP: Change the $Pr$ to $\Pr$.}

{We are interested in maximizing the probability of decoding of a packet irrespective of its source. The union over $r$ in (\ref{eq:main_obj}) captures that if any of the repetitions $r=1,...,R$ is decoded, the packet $n$ is considered decoded. 
The union over $b$ captures that a packet repetition is decoded if at any of the locations $b$ there is a BS synchronized to band $\beta(n,r)$ (i.e. $\left[\mathbf{X}\right]_{b,\beta(n,r)} = 1$) that is able to decode it. A BS can decode a repetition at location $b$ if the SINR of the packet repetition at the location $b$, $\text{SINR}_{b,\beta(n,r)}(n,r)$, exceeds the decoding threshold $\tau$. }
%{The decoding of individual repetitions of a packet will be correlated since they have the same source, therefore they may experience similar interference and shadowing. This is especially true when the repetitions happen to occur over the same multiplexing band. }
%\pp{PP: The equation below looks quite entangled, you need to explain what the individual parts are. Furthermore, do we assume anything about the temporal dynamics of the transmissions, i.e. the probability that a certain IoT device is active? Explain that $r$ is the number of repetitions. We aslo need to mention that when $r>1$ there is correlation among the SINRs achieved in interfering transmissions if they all use the same frequency.} 

The optimization problem can be stated as:

\begin{subequations}
\begin{align} \label{eq:P1}
\max_{\mathbf{X}} & \quad P_{\text{PDP}}(\mathbf{X}) \tag{P.1.1} \\
\textrm{s.t.} & \quad \left[\begin{array}{c}
\mathbf{1}_{B} \\
\mathbf{0}_{C}
\end{array}\right] \leq \mathbf{X}\mathbf{1}_{M} \leq \left[\begin{array}{c}
\mathbf{1}_{B} \\
\mathbf{1}_{C}
\end{array}\right], \tag{P1.2} \label{P1.2} \\
&\left[\begin{array}{cc}
\mathbf{0}_{B}^{T} & \mathbf{1}_{C}^{T}\end{array}\right]\mathbf{X}\mathbf{1}_{M}=\Delta B, \tag{P1.3} \label{P1.3} \\
& \quad \mathbf{X}\in \mathbb{Z}_2^{(B+C) \times M} \tag{P1.4} \label{P1.4} 
\end{align}
\end{subequations}
In the above, $\mathbf{1}_{B}$ is a vector of ones of size $B \times 1$ and $\mathbf{0}_{C}$ is a vector of zeros of size $C \times 1$. The constraint (\ref{P1.2}) ensures that BSs $b\in\{1,\dots,B\}$ that are already installed, remain at their location and are assigned to listen to a multiplexing band and none of the BSs are assigned to more than one band. The constraint (\ref{P1.3}) ensures that $\Delta B$ BSs are placed to one of the $C$ candidate locations and assigned to one of the $M$ multiplexing bands. The constraint (\ref{P1.4}) constrains the assignment variable $\mathbf{X}$ to have entries limited to $\{0,1\}$.

In order to be able to solve this problem, we would have to know what the function  $P_{\text{PDP}}(\mathbf{X})$ is in terms of the assignment variable $\mathbf{X}$. 
Learning or estimating this function would be non-trivial since the number of possible $\mathbf{X}$ is ${C \choose {\Delta B}} M^{B + \Delta B}$. 

\subsection{Suboptimal solution to P1}
% The relaxation is that this is the lower bound
%probability of a transmission irrespective of the source,

%$
%    P_{\text{TDP}}(\mathbf{X})=Pr\left(\bigcup_{b}\left[\mathbf{X}\right]_{b,\beta(r)}\text{SINR}_{b,\beta(r)}(r)>\tau\right)
%    \label{eq:new_obj}
%$, 

%which is a lower bound on $P_{\text{PDP}}(\mathbf{X})$. 

To relax the optimization problem (P1), we seek to optimize the probability of decoding of a repetition of a packet, 
$
    P_{\text{TDP}}(\mathbf{X})=\Pr\left(\bigcup_{b}\left[\mathbf{X}\right]_{b,\beta(r)}\text{SINR}_{b,\beta(r)}(r)>\tau\right)
$, 
which is a lower bound on $P_{\text{PDP}}(\mathbf{X})$. $P_{\text{TDP}}(\mathbf{X})$ is a lower bound on $P_{\text{PDP}}(\mathbf{X})$  since a packet is decoded if any of its $R$ repetitions is decoded, therefore $P_{\text{PDP}}(\mathbf{X})$ is always larger than $P_{\text{TDP}}(\mathbf{X})$. As we will demonstrate in this section, working with the expression for $P_{\text{TDP}}(\mathbf{X})$ makes the analysis more tractable and allows us to arrive at a simpler optimization problem. The relaxed problem (P1) is:
% \begin{equation}
% \begin{aligned}
% \max_{\mathbf{X}} & \quad P_{\text{TDP}}(\mathbf{X})  \\
% \textrm{s.t.} & \quad \left[\begin{array}{c}
% \mathbf{1}_{B} \\
% \mathbf{0}_{C}
% \end{array}\right] \leq \mathbf{X}\mathbf{1}_{M} \leq \left[\begin{array}{c}
% \mathbf{1}_{B} \\
% \mathbf{1}_{C}
% \end{array}\right],  \\
% &\left[\begin{array}{cc}
% \mathbf{0}_{B}^{T} & \mathbf{1}_{C}^{T}\end{array}\right]\mathbf{X}\mathbf{1}_{M}=\Delta B, \\
% & \quad \mathbf{X}\in \mathbb{Z}_2^{(B+C) \times M} 
% \end{aligned}
% \tag{P2}
% \label{P2}
% \end{equation}
% Shortened the formulation
\begin{equation}
\begin{aligned}
\max_{\mathbf{X}} & \quad P_{\text{TDP}}(\mathbf{X})  \\
\textrm{s.t.} & \quad \text{(\ref{P1.2}), (\ref{P1.3}),  (\ref{P1.4})}
\end{aligned}
\tag{P2}
\label{P2}
\end{equation}
{
Note that (\ref{P2}) is equivalent to (P1) when the number of repetitions is $R = 1$. 
%This also means that the assignment obtained by solving (\ref{P2}) is independent of $R$.
}
\begin{proposition}{}
\label{prop:suboptimal}
For a given $\tau$, a suboptimal solution to \ref{P2} can be obtained by solving the following convex integer optimization problem:   
\begin{equation}
\begin{aligned}
\max_{\mathbf{X}} \quad & \sum_{m}\bigg(\sum_{b}[\mathbf{X}]_{b,m}\mathbb{E}_r\{A_{b,m}(r)\}-\\
&\sum_{b<v}[\mathbf{X}]_{b,m}[\mathbf{X}]_{v,m}\mathbb{E}_r\{A_{b,m}(r)A_{v,m}(r)\}\bigg)\\
\textrm{s.t.} & \quad \text{(\ref{P1.2}), (\ref{P1.3}),  (\ref{P1.4})}
\end{aligned}
\tag{P3}
\label{P3}
\end{equation}
where $A_{b,m}(r) = \mathbf{1_{(\text{SINR}_{b,\beta(n,r)}(n,r)>\tau)}}$. The objective function is a concave function and is also a lower bound on the objective function in (\ref{P3}). Therefore, the objective function in (\ref{P3}) is a lower bound on the objective function in (P1).
\end{proposition}
\begin{IEEEproof}
The proof is given in Appendix \ref{appendix:suboptimal}.
\end{IEEEproof}

{\subsection{Practical solution of the proposed optimization problem}}
As shown by our results in the latter sections, approximating the problem (P1) by (\ref{P3}) proves to be a suitable relaxation for finding the BS assignment and placement that will result in maximizing the PDP in a realistic environment.

{While the solution given by (\ref{P3}) can be used to maximize the lower bound on packet decoding probability, in order to implement it, we need to know $\mathbb{E}_r\{A_{b,m}\}~\forall b$, which are the ADPs at each of the locations in the set $\{1,\dots,B+C\}$. We also need the knowledge of $\mathbb{E}_r\{A_{b,m}(r)A_{v,m}(r)\} ~ \forall b,v$, which are the JDPs between all pairs of locations in the set $\{1,\dots,B+C\}$. {Additionally, even with the knowledge of ADPs and JDPs the problem may be difficult to solve due to integer constraints on the optimization variables.}

{While general integer optimization problems are known to be NP-hard, the optimization problem (\ref{P3}) stands out in that its objective function is convex and its constraints are linear. It can be shown that there exists an algorithm that can solve a convex integer problem in polynomial time given a fixed number of variables \cite[p.~574]{junger200950}. However, the worst-case computational complexity asymptotically grows exponentially with the number of variables. In our case, the number of variables is $M\times(B+C)$. Therefore, even though this problem is solvable it may not be tractable. In our solution we use a branch-and-bound algorithm to solve this problem and place a $T_{sol}$ threshold on the solver CPU time as a stopping criterion. $T_{sol}$ was selected such that in the majority of cases the optimal solution was found on our machine. To guarantee tractability, relaxations of the problem (\ref{P3}) could be pursued to lower computational complexity, however that is beyond of the scope of this paper}.  

Therefore, the main challenge that we will focus on is to accurately and efficiently estimate the ADPs $\mathbb{E}_r\{A_{b,m}\}~\forall b$ and JDPs $\mathbb{E}_r\{A_{b,m}(r)A_{v,m}(r)\} ~ \forall b,v$. Due to this critical aspect of our problem, we will propose two alternative algorithms for BS placement and band assignment:
% \begin{enumerate}
%     \item 
%     %\emph{Model-based (MOD) placement and assignment approach}: In this approach, we will use models of ADP and JDP to predict $\mathbb{E}_r\{A_{b,m}\}~\forall b$ and $\mathbb{E}_r\{A_{b,m}(r)A_{v,m}(r)\} ~ \forall b,v$. 
%     \emph{Model-based (MOD) placement and assignment approach}: In this approach, we will use models of ADP and JDP to predict $\mathbb{E}_r\{A_{b,m}\}~\forall b$ and $\mathbb{E}_r\{A_{b,m}(r)A_{v,m}(r)\} ~ \forall b,v$.
%     This approach necessitates development of suitable models of ADP and JDP and estimating the parameters of these models to make the correct predictions on ADP and JDP. 
%      \item \emph{Measurement-based (MEAS) placement and assignment approach}: In this approach, we directly measure $\mathbb{E}_r\{A_{b,m}\}~\forall b$ and $\mathbb{E}_r\{A_{b,m}(r)A_{v,m}(r)\} ~ \forall b,v$ through an optimized training procedure during the operation of the network. 
% \end{enumerate}
\begin{enumerate}
     \item \emph{Measurement-based (MEAS) placement and assignment approach}: In this approach, we directly measure $\mathbb{E}_r\{A_{b,m}\}~\forall b$ and $\mathbb{E}_r\{A_{b,m}(r)A_{v,m}(r)\} ~ \forall b,v$ through an optimized training procedure during the operation of the network. However, the complexity of this training procedure scales with the number of candidate locations since the number of JDP parameters to be estimated is ${B+C}\choose{2}$. For this reason, we have developed an alternative, more practical approach.
    \item \emph{Model-based (MOD) placement and assignment approach}: In this approach, we will use models of ADP and JDP to predict $\mathbb{E}_r\{A_{b,m}\}~\forall b$ and $\mathbb{E}_r\{A_{b,m}(r)A_{v,m}(r)\} ~ \forall b,v$.
    This approach necessitates development of suitable models of ADP and JDP and estimating the parameters of these models to make the correct predictions on ADP and JDP. 

\end{enumerate}
% TODO List:
% - Derive the expression for PDP in terms of TDP
% - Move the channel model into the section III

%The MOD approach has a lower training complexity than the MEAS approach, however it requires the environment to meet certain criteria that the models are based on. 
%On the other hand, the MEAS approach has very loose criteria that the environment should match.

\section{Model-based placement and assignment approach}
\label{sec:model-based-approach}

In this section, we describe the proposed model-based approach. In Sec. \ref{sec:model-adp}, we will develop a model of ADP as a function of a set of parameters $\psi_{\ADP, m}$ dependent on band $m$, denoted as  $\ADPM$ (Theorem \ref{theorem:adp-model}). 
In Sec. \ref{sec:model-jdp}}, we will develop a model of JDP as a function of a set of parameters, $\JDPPARAM$, 
and the separation between receiving locations $b$ and $v$, $d_{b,v}$, denoted as $\JDPM$ (Theorem \ref{theorem:model-jdp}). 
Then, the models of ADP and JDP can be applied in a real environment by estimating the parameters $\psi_{\ADP,m}$ and $\psi_{\JDP,m}$ and using the estimated parameters to predict  ADPs and JDPs as: 
\begin{equation}
\mathbb{E}_r\{A_{b,m}(r)\} = \ADPM~\forall b,   
\end{equation}
\begin{multline}
\mathbb{E}_r\{A_{b,m}(r)A_{v,m}(r)\} =  \\ \JDPM ~  \forall b,v
\end{multline}
%$$

Hence,  (\ref{P3}) can be solved using the predicted ADPs and JDPs. 

\subsection{Stochastic geometry preliminaries}
\label{sec:prelim}
In this section, we introduce some fundamental concepts from stochastic geometry that will assist us in development of the model of the ADP, $\ADPM$, and the model of the JDP, $\JDPM$. 

%A wireless network can be viewed as a collection of nodes, including transmitters and receivers, whose quantity and locations can be viewed as realizations of random processes. The performance metrics of the wireless network such as the probability of a successful transmission or average interference power per receiver are dependent on the number and the arrangement of the nodes in the network. Assuming that the underlying model of the number and placement of nodes is ergodic, we can model the expected performance metrics over many realizations of the number of nodes and their placement using stochastic geometry. Stochastic geometry is an area of mathematics that uses applied probability to predict the averages of random phenomena on the plane or in higher dimensions. Initially, its development was stimulated by applications to biology, astronomy and material sciences, but in the past century it has proven effective in the context of communications networks \cite{baccelli2009stochastic}.

Stochastic geometry is linked to spatial point processes. A point process (PP) is a countable random collection of points that reside in some measure space, usually the Euclidean space $\mathbb{R}^d$ \cite{haenggi2012stochastic}. Formally, a PP is a countable random set
$\Phi = \{x_1,x_2,...\} \in \mathbb{R}^d$ consisting of random variables $x_i \in \mathbb{R}^d$ as its elements.

A PP model that is often used to model the topologies of wireless networks is the Homogeneous Poisson Point Process (HPPP). A HPPP with density $\lambda$ is a PP in $\mathbb{R}^d$ such that for every compact set $\mathcal{A}$, the number of points has a Poisson distribution with mean $\lambda |\mathcal{A}|$, where $|\cdot|$ is the Lebesgue
measure in $d$ dimensions, and if two compact sets $\mathcal{A}_1$ and $\mathcal{A}_2$ are disjoint, then the number of points in each is independent of one other. 

One important result for the {analysis} of HPPP that we will utilize is the probability generating functional (PFGL) of a HPPP. Let $\mathcal{V}$ be the family of measurable functions, then for a $v(x) \in \mathcal{V}$ and a HPPP with density $\lambda$ \cite[p.~150]{haenggi2012stochastic}:
\begin{equation}
    \mathbb{E}_{\Phi}\left(\prod_{x\in\Phi} v(x) \right) =  \exp\left(-\int_{\mathbb{R}^d}[1-v(x)]\lambda dx\right).
\end{equation}
With these basic concepts covered, we will explain how stochastic geometry can be used to model our problem.  

\subsection{Stochastic modeling of the IoT network}

{In this section, we describe a way of modeling the activity of IoT and incumbent devices as random processes. In our modeling, we preserve the assumptions about the IoT network and the {environment, stated in Sec. \ref{sec:network-model}}. 
%Furthermore, we describe our model for the SINR of packet transmissions. 
By applying
theorems from stochastic geometry on these random processes, we will develop models of the
ADP and JDP in the {later} sub-sections.}}

% \begin{figure}[t]
%     \centering
%     \includegraphics[width=0.6\linewidth]{figures/is_this_a_ppp.eps}
%     \caption{Simulated PMF of $|\Phi|$ and the PMF of a Poisson distribution with density $Dp$, where $p$ is calculated using Eq. \ref{eq:what_is_p}. To obtain the graph, a disk shaped UNB IoT netwo             zzzzzzzzzzzzzzzzzzzzzzzzzzzzzzzzzzrk with a radius of $10$ km and D = 3000. Other parameters were $M=3$, $N=3$, $R=3$, $T = 0.01$ s, $w=1$ kHz, $W=200$ kHz.}
%     \label{fig:is_this_a_ppp}
% \end{figure}

\subsubsection{Modeling the topology of active devices as a HPPP}

While we assume that the number and the position of the nodes in the network is fixed, given that the IoT and incumbent devices access the medium using the ALOHA protocol, the set of devices that are transmitting at some time $\Tilde{t}$ and frequency $\Tilde{f}$ in $|\mathcal{A}|$ is random.  
We will argue that that the distribution of both IoT and incumbent devices of active devices at some time $\Tilde{t}$ and frequency $\Tilde{f}$ can be modeled as a HPPP and can then be analyzed using the tools from stochastic geometry. Furthermore, it is important to emphasize that the analysis of networks using stochastic geometry is in most cases applied to finding the average of some random quantity over many realizations of wireless networks sampled from some spatial point process. In our case, we are interested in a deterministic topology of devices but as we will argue in this section due to the random access nature of IoT and incumbent devices, their activity from one time slot to the next can be approximately modeled as a HPPP. 
%\pp{PP: We have to relate this to the temporal behavior of the random ALOHA-like transmissions.}

Let us focus on the IoT devices first, since our model for the incumbent network is similar in all aspects except that incumbents are operating over a single multiplexing band. {Since we assume an ALOHA-like protocol and Markovian packet arrivals}, the probability of an IoT device being active at some time $\Tilde{t}$ and frequency $\Tilde{f}$ is $p$, where 
\begin{equation}
p = 2\frac{NRT}{1~h} \times 2 \frac{w}{MW}.    
\label{eq:what_is_p}
\end{equation}
Let the set of active IoT devices at time $\Tilde{t}$ and frequency $\Tilde{f}$ be $\Phi$. Then the size of $\Phi$, $|\Phi|$, is {given by a Binomial distributions}, $\Pr(|\Phi|) = {D\choose |\Phi|}p^{|\Phi|}(1-p)^{(D-|\Phi|)}$. Since $D$ is large and $p \ll 1$, we can approximate the Binomial distribution model of $|\Phi|$ by a Poisson distribution with a parameter $Dp$, $\Pr(|\Phi|) = \frac{(Dp)^{|\Phi|} \exp(-Dp)}{|\Phi|!}$. According to a rule of thumb, this approximation is good if $D$ is large and $p$ is small, and $Dp$ is not large \cite{steele1994cam}.
Furthermore, the devices in $\Phi$ are uniformly distributed across $\mathcal{A}$, therefore the set of active devices $\Phi$ can be approximated to be a realization of an HPPP with density $\lambda = \frac{Dp}{|\mathcal{A}|}$. 

%In Fig. \ref{fig:is_this_a_ppp}, we show how the simulated PMF of $|\Phi|$ and the PMF of a Poisson distribution with density $Dp$, where $p$ is calculated using Eq. \ref{eq:what_is_p} closely match one another. 

Using a similar reasoning, we can approximately model the set of active incumbent nodes ${\Phi}_I$ at time $\Tilde{t}$ and frequency $\Tilde{f}$ as a HPPP with density ${\lambda}_{I,m} = \frac{{D}_{I,m}{p}_{I,m}}{|\mathcal{A}|}$, where the density of the incumbent devices depends on the multiplexing band $m$.

Note that the activity of incumbent and IoT devices could be modelled slightly more accurately using a homogenous binomial point process (HBPP), where the number of points in a closed set is modeled using a Binomial distribution. However, we choose to use the HPPP over the HBPP since it allows us to arrive at some general expressions for models of ADP and JDP.

\subsubsection{Modelling the SINR of transmissions}

Let the SINR of some transmission $(n,r)$ from IoT device $i$ to BS $b$ over multiplexing band $m$ be $\text{SINR}_{i,b,m}(n,r)$.
% , and it can be expressed (in dB) as 
% $\gamma_{b,m}(n) = P_b(n) - (P^N(n) + P^I_{b,m}(n)),$ 
% where $P^N(n)$ is the noise during transmission $n$ and $P^I_{b,m}(n)$ is the interference power during transmission $n$ at BS $b$ across multiplexing band $m$. $P^N(n)$ is an i.i.d. random variable and $P^I_{b,m}(n)$ is a random variable that is i.i.d. across $n$ but may be correlated across $b$ and $m$. For example, two nearby BSs or two neighboring multiplexing bands may experience interference from similar devices. 
%{The algorithm does not rely on any other assumptions on the distributions of $P^N(n)$ and $P^I_{b,m}(n)$.}
%Since the bandwidth of the multiplexing band is relatively small (on the order of hundreds of kHz), we assume that the probability distribution of $\gamma_j[n]$ is not dependent on the multiplexing band on which the transmission takes place. This assumption relies  does not change significantly across multiplexing bands and that the propagation channel behaves similarly across bands.
A transmission $(n,r)$ is considered to be successfully decoded if $\text{SINR}_{i,b,m}(n,r)$ at any of BS $b$ listening to the band $\beta(n,r)$ exceeds a threshold $\tau$. In practice, $\tau$ can be the minimum SINR required to achieve a certain bit error rate (BER) performance.
The value of $\tau$ depends on specific coding, modulation, and detection schemes being employed. For example, in the Sigfox technical documentation it is stated that a good detection probability can be achieved if the SINR exceeds 8 dB \cite{sigfox2017sigfox}. The $\text{SINR}_{i,b,m}$ from an IoT device $i$ is modeled as 
\begin{equation}
\text{SINR}_{i,b,m}=\frac{h_{i}p_{i,b}^{-\alpha}}{\hat{P}_{N}+
\sum_{j\in\Phi}f_{j}p_{j,b}^{-\alpha}+\sum_{j'\in{\Phi}_{I,m}} \hat{P}_I f_{j'}p_{j',b}^{-\alpha}}
\end{equation}
where $p_{i,b}$ is the distance of the device $i$ to BS $b$, $p_{j,b}$ is the distance of an interfering IoT device $j$, and $p_{j',b}$ is the distance of an incumbent device $j'$. $\alpha$ is the path-loss exponent and $h_i$, $f_j$ and $g_{j'}$ are fading gains, modeled as Exp$(1)$. {$\hat{P}_N$ and $\hat{P}_I$ are noise power and incumbent device power expressed as a fraction of the IoT device transmission power. In order for the path loss function $p^{-\alpha}$ to be integrable ($\int_0^{\infty} p \times p^{-\alpha} dp < \infty$)}, we assume that $\alpha > 2$.

%Since we assume that all devices in the IoT network make transmissions at an equal rate, to model the SINR of a typical user $i$, $\text{SINR}_{i,b,m}$, is equivalent to modeling the SINR of a typical transmission, $\text{SINR}_{b,m}(n,r)$, which is why from now on we will use $\text{SINR}_{b,m}$ to refer to both the SINR of a typical user or of a typical transmission.

\title{Optimal band assignment for UNB LPWA networks with multiband access}

%\pp{PP: We need a comment whether this problem is optimized in a centralized or distributed manner as well as who needs to know what in order to carry out the optimization.}

\subsection{Modeling the ADP}

\newtheorem{theorem}{Theorem}
\label{sec:model-adp}
We first derive the expression for the probability of  decoding a transmission from a device $i$ at a receiver $b$, on band $m$, denoted as $\Pr\left(\text{SINR}_{i,b,m}\geq\tau~|~p_{i,b}\right)$. Then we use $\Pr\left(\text{SINR}_{i,b,m}\geq\tau~|~p_{i,b}\right)$ to derive a model of the ADP,  $\Pr(\text{SINR}_{b,m}\geq\tau)$.
\begin{theorem}
\label{theorem:ccdf-sinr}
The  complementary cumulative distribution (CCDF) of $\text{SINR}_{i,b,m}$ when noise power is negligible ($\hat{P}_N\rightarrow0$) is: 
\begin{equation}
\Pr(\text{SINR}_{i,b,m}\geq\tau\mid p_{i,b}) 
= \exp\left(-\epsilon_m\tau^{\frac{2}{\alpha}}p_{i,b}^{2}\right),
\label{eq:sinr-ccdf}
\end{equation}
where $\epsilon_m = \pi\left(\lambda+\hat{P}_{I}^{\frac{2}{\alpha}}{\lambda}_{I,m}\right)\frac{2\pi/\alpha}{\sin(2\pi/\alpha)}$. 
%The ADP decays exponentially as a function of three terms: $p_{i,b}^{2}$, $\tau^{\frac{2}{\alpha}}$ and $\epsilon_m$. The term $p_{i,b}^{2}$ tells us that the ADP decays exponentially as a function of squared distance to the BS. $\tau^{\frac{2}{\alpha}}$ indicates that ADP decays as the decoding threshold $\tau$ increases but the extent of it is dependent on the path loss exponent $\alpha$. 
The term $\epsilon_m$ is proportional to the interference in the network, since it depends on the density of incumbent and IoT users. {Note that as $\alpha$ increases, the term $\hat{P}_{I}^{\frac{2}{\alpha}}{\lambda}_{I,m}$ approaches ${\lambda}_{I,m}$, in cases when either $\hat{P}_{I}>1$ (the incumbent power is stronger than IoT device power) or when $\hat{P}_{I}\leq1$, which indicates that as $\alpha$ increases, the power of incumbent devices relative to IoT devices becomes less important, and only the density matters.} This is sensible since when $\alpha$ is large, the signal power from distant devices decays significantly and only the activity of devices close to the BS is important. However, if an incumbent device close to the BS is active during a transmission its transmit power is not relevant as it will likely have a higher power  than the IoT transmission anyway due its proximity to the BS.  
\end{theorem}
\begin{IEEEproof}
The proof is shown in Appendix \ref{proof:ccdf-sinr}.
\end{IEEEproof}

%We will now evaluate the ADP irrespective of the receiver for a finite network $\mathcal{A}$ around the receiver $b$.
We will now evaluate the ADP {across all transmitters} for a finite network $\mathcal{A}$ around the receiver $b$.
\begin{theorem}
The ADP irrespective of the source is
\begin{equation}
   \Pr\left(\text{SINR}_{b,m}\geq\tau \right)  =  \int_{\mathcal{A}} \exp\left(-\psi_mp_{i,b}^{2}\right)\frac{p_{i}}{|\mathcal{A}|}dp_{i}\theta_{i}, 
   \label{eq:adp-model}
\end{equation}

\begin{figure}[t]
    \centering
    \centering
    \includegraphics[width=0.8\linewidth]{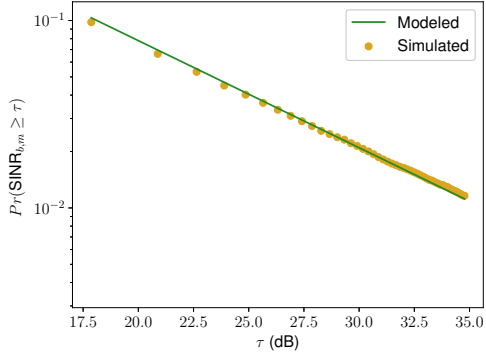}
    \caption{Simulated and modeled ADP as a function of the decoding threshold $\tau$. The modeled results are obtained using (\ref{eq:adp-model}).
    We have simulated a disk shaped UNB IoT network with the following parameters: $300$km radius, $D = D_I = 15E6$, $M=1$, $N=3$, $R=3$, $w=600$Hz, $w_I=200$kHz, $W=200$kHz, $T=(208/w)$s, and $T_I=(2080/w_I)$s.}
    \label{fig:ccdf-sinr}
\end{figure}

\begin{figure}[t]
    \vspace{-10pt}
    \centering
    \includegraphics[width=0.75\linewidth]{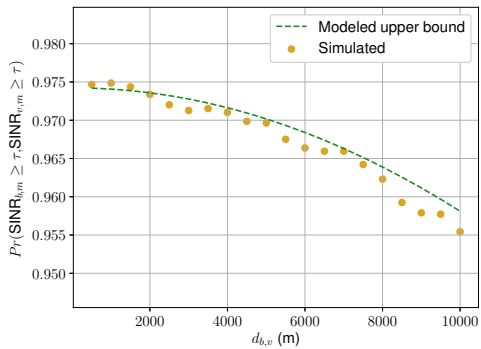}
    \caption{Simulated and modeled JDP as a function of the separation between BS, $d_{b,v}$. The modeled results are obtained using the derived expression in \ref{eq:jdp-model}.
    The simulated results are obtained by measuring the average rate of transmissions decoded by both BS $b$ and $v$. 
    The graph is obtained in the same simulation environment as the one described in Fig. \ref{fig:ccdf-sinr}. Additionally, we set $\tau=0$ dB.} 
    \label{fig:jdp-sinr}
\end{figure}

where $|\mathcal{A}|$ and $(p_{i},\theta_{i})$ are the coordinates of the device $i$ with respect to the origin.
Let $(p_{b},\theta_{b})$ be the polar coordinates of the receiving location, then $p_{i,b}^2=p_{i}^2+p_{b}^2-2p_ip_b\cos(\theta_i-\theta_b)$. Moreover, $\psi_m=\epsilon_m\tau^{\frac{2}{\alpha}}$, and hence $\ADPPARAM=\{\psi_m, \mathcal{A}\}$. 
The unit of $\psi_m$ is m$^{-2}$.
\label{theorem:adp-model}
\end{theorem}
\begin{IEEEproof}
The expression is obtained by taking the expectation over $p_i$ and $\theta_i$ in the expression in (\ref{eq:sinr-ccdf}).
The joint PDF of $p_i$ and $\theta_i$ is $\frac{p_{i}dp_{i}\theta_{i}}{|\mathcal{A}|}$.  
\end{IEEEproof}
To the best of our knowledge, the integral in (\ref{eq:adp-model}) cannot be simplified to a closed-form expression, hence when evaluating the ADP we will use  numerical integration.
%We will make the approximation that $f(p_{i,b})$ is equal for all $b$, hence the average decoding rate $Pr\left(\text{SINR}_{b,m}\geq\tau \right)$ is dependent only on $m$. The approximation that $f(p_{i,b})$ is equal for all $b$ is valid if the size of $\mathcal{A}$ is much larger than the separation between BSs in the network. 

In Fig. \ref{fig:ccdf-sinr}, we show the simulated ADP as a function of the decoding threshold $\tau$ for a disk-shaped $\mathcal{A}$ centered around the receiver $b$. The ADP modeled using (\ref{eq:adp-model}) closely matches the simulated results which supports our model of the IoT network as a HPPP and the model in (\ref{eq:adp-model}).

\subsection{Modeling the JDP}
\label{sec:model-jdp}
\begin{figure}[t]
    \centering
    \includegraphics[width=0.6\linewidth]{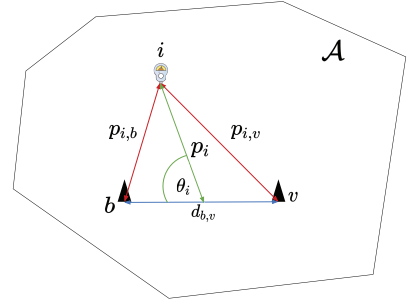}
    %\vspace{-5pt}
    \caption{Illustration of the example used to model the joint decoding probability between two BSs.}
    \label{fig:dipole}
\end{figure}

Here, we derive the expression for the JDP, denoted as $\Pr(\text{SINR}_{i,b,m}\geq\tau,\text{SINR}_{i,v,m}\geq\tau)$. First, we will derive the expression for the upper bound on the JDP for a given transmitter $i$, denoted as $\Pr(\text{SINR}_{i,b,m}\geq\tau,\text{SINR}_{i,v,m}\geq\tau~|~i)$. We assume that the BSs $b$ and $v$ are located at the origin of the polar coordinate system as shown in Fig. \ref{fig:dipole} with coordinates $(\frac{d_{b,v}}{2}, \frac{\pi}{2})$ and $(\frac{d_{b,v}}{2}, -\frac{\pi}{2})$, respectively. 
\begin{theorem}
\label{theorem:joint-pdp}
The  upper bound on conditional JDP  for a given transmitter $i$ is: 
\begin{multline}
\Pr(\text{SINR}_{i,b,m}\geq\tau,\text{SINR}_{i,v,m}\geq\tau~|~i) \leq \\
\exp\left(-\frac{1}{2}\epsilon_m\tau^{\frac{2}{\alpha}}d_{b,v}^{2}\right) \times 
\exp\left(\left(-\frac{2}{\alpha}-1\right)\epsilon_m\tau^{\frac{2}{\alpha}}p_i^{2}\right)
\label{eq:jdp}
\end{multline}
where $p_i$ is the distance of a device $i$ to the origin of the coordinate system. Note that the validity of this model depends on the approximation that the midpoint between the two BSs is located at the origin, therefore it can only serve as an approximation for BS pairs that are located near the edge.
\end{theorem}
\begin{IEEEproof}
The proof is shown in Appendix \ref{proof:joint-pdp}.
\end{IEEEproof}

\begin{theorem}
The JDP irrespective of the source is:
\begin{equation}
\Pr\left(\text{SINR}_{b,m}\geq\tau,\text{SINR}_{v,m}\geq\tau\right) = 
\Psi_m\exp\left(-\frac{1}{2}\psi_m d_{b,v}^{2}\right)
\label{eq:jdp-model}
\end{equation}
where $\Psi_m = \int_{\mathcal{A}}\exp\left(\left(-\frac{2}{\alpha}-1\right)\epsilon_m\tau^{\frac{2}{\alpha}}p_i^{2}\right)$. Then, $\JDPPARAM=\{\psi_m,\Psi_m\}$. This theorem allows us to model JDP between two BSs as a function of their separation, $d_{b,v}$.
\label{theorem:model-jdp}
\end{theorem}

\begin{IEEEproof}
The proof is obtained by taking the expectation with respect to $p_i$ of the expression in (\ref{eq:jdp}). 
\end{IEEEproof}

In Fig. \ref{fig:jdp-sinr}, we show the simulated JDP for two BSs, $b$ and $v$, whose separation $d_{b,v}$ is varied. 
The JDP modeled using Theorem \ref{theorem:model-jdp} is an upper bound to the simulated results.

\subsection{Summary of the proposed algorithm}
In Algorithm \ref{alg:assign}, we summarize the MOD approach for assignment and placement assuming that $\JDPPARAM$, $\ADPPARAM$ and positions of the BSs are known. The algorithm is based on predicting ADP and JDP using Theorem \ref{theorem:adp-model} and Theorem \ref{theorem:model-jdp}, respectively. The predicted ADPs and JDPs are used in (\ref{P3}) to solve for optimal placement and frequency assignment of BSs. 

In Sec. \ref{sec:training}, we describe the method for estimation of parameters of $\JDPPARAM$ and $\ADPPARAM$ during network operation. 
\begin{algorithm}[t!]
\DontPrintSemicolon
%\SetAlgoLined
\KwIn{$\{\psi_m,\Psi_m\}~\forall m$,~$(p_b,\theta_b)~\forall b$,~$\mathcal{A}$}
\KwOut{Assignment $\mathbf{X}$}
\For{$b \in \{1,\dots,B+C\}$}{

\% Estimate the ADP

$\mathbb{E}_r\{A_{b,m}(r)\} \leftarrow \int_{\mathcal{A}} \exp\left(-\psi_mp_{i,b}^{2}\right)\frac{p_{i}}{|\mathcal{A}|}dp_{i}\theta_{i}~\forall m$

    \For{$v \in \{1,\dots,b\}$}{
    $d_{b,v} \leftarrow \sqrt{p_b^2+p_v^2-2p_bp_v\cos(\theta_b-\theta_v)}$
    
    \% Estimate the JDP
    
    $\mathbb{E}_r\{A_{b,m}(r)A_{v,m}(r)\} \leftarrow \Psi_m\exp\left(-\frac{1}{2}\psi_m d_{b,v}^{2}\right)~\forall m$
    
    }
}

Solve for $\mathbf{X}$ in (\ref{P3})

%$S_{b,m} \leftarrow \frac{1}{\sum_{l \in  \mathcal{L}}\mathbf{1}(z_b[l]=m)}\sum_{l \in  \mathcal{L}}y_{b,m}[l]~\forall~b,m$ \;
%$R_{b,k,m} \leftarrow \frac{1}{\sum_{l \in  \mathcal{L}}\mathbf{1}(z_b[l],z_k[l]=m)}\sum_{l \in\mathcal{L}}y_{b,m}[l]y_{k,m}[l]~\forall~b,k,m$\;

\caption{Model-based placement and frequency assignment algorithm}
\label{alg:assign}
\end{algorithm}

\section{Learning approach for model-based assignment}
\label{sec:training}

\subsection{Parameter estimation}
\label{sec:training-parameter-estimation}
As shown in Algorithm \ref{alg:assign}, to perform the assignment, our approach requires the information about $\Psi_m$ and $\psi_m$ for all $m$, locations of current BSs and candidate locations and the contour 
%\pp{PP: Note that $\mathcal{A}$ has been introduced before, without being defined precisely what it is. I have mentioned the term ``area'' before, but here ``contour'' is mentioned, please specify earlier.} \red{EK: Ghaith had a similar comment. I added a more precise definition of $\mathcal{A}$ in Sec. III.C}
of the {area} $\mathcal{A}$. It is assumed that the contour of the {area} $\mathcal{A}$ is known by the network operator network, without assuming that the locations of the IoT devices are known or their number. We will now discuss how to estimate parameters $\psi_m$ and $\Psi_m$ for all $m$ during the operation of the network.

We will use {estimates} of JDP between BSs in $\mathcal{A}$ to estimate $\psi_m$ and $\Psi_m$ for all $m$. 
%Let us assume that there are $S_m$ measurements of JDP collected on each band and let the set of JDP measurements on each band be $\mathcal{S}_m$.
Furthermore, we will denote the estimates JDP between two locations $b$ and $v$ as $\tilde{\mathbb{E}}_r\{A_{b,m}(r)A_{v,m}(r)\}$.
{Let us denote the set of JDP estimates on each band as $\mathcal{S}_m$, where $|\mathcal{S}_m| = S_m$.}
To estimate $\psi_m$ and $\Psi_m$ for all $m$, we will use the least squares criterion between the measured JDP and the fitted model. Parameter estimation is then a non-linear least squares (NLSQ) optimzation problem:
\begin{equation}
\begin{aligned}
\min_{\psi_m, \Psi_m} & \sum_{m \in\{1,\dots,M\} } \sum_{(b,v) \in \mathcal{S}_m }|\tilde{\mathbb{E}}_r\{A_{b,m}(r)A_{v,m}(r)\}- \\
& \Psi_m\exp{\left(-\psi_m d_{b,v}^2\right)}|^2 
\end{aligned}
\tag{P4}
\label{P4}
\end{equation}
which can be solved using the Levenberg–Marquardt algorithm \cite{more1978levenberg} for all $m$.

\begin{figure}[t]
    \centering
    \includegraphics[width=0.8\linewidth]{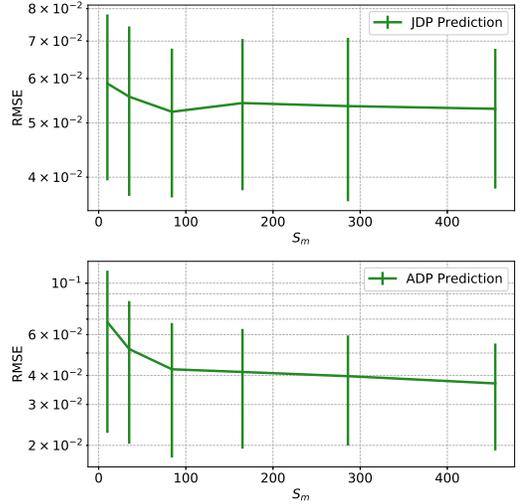}
    %\vspace{-20pt}
    \caption{Root mean square error (RMSE) in prediction of ADP and JDP at a random location in $\mathcal{A}$ as a function of $S_m$ used to estimate the ADP and JDP model parameters. The graph is obtained in the same simulation environment as the one described in Fig. \ref{fig:ccdf-sinr}}
    \label{fig:interpolation}
\end{figure}

%After we have estimated $\psi_m, \Psi_m$  $\forall m$,  ADPs $\mathbb{E}_r\{A_{b,m}\}~\forall b \in \{1,\dots,B+C\}$ are predicted using the Theorem \ref{theorem:adp-model}. Similarly, JDPs $\mathbb{E}_r\{A_{b,m}(r)A_{v,m}(r)\} ~ \forall b,v \in \{1,\dots,B+C\}$ are predicted using Theorem \ref{theorem:model-jdp}. 
The root mean square error (RMSE) of ADP and JDP prediction as a function of $S_m$ is shown in Fig. \ref{fig:interpolation}. Based on the simulation results, we observe that the accuracy of prediction stops significantly decreasing past $S_m = 10$ and saturates past $S_m = 20$. 

\subsection{Training procedure for estimation of $\ADPPARAM$ and $\JDPPARAM$}
\label{sec:mod-training-procedure}
In this section, we describe the training procedure for collection of measurements of JDP. The training episode is executed to keep up with changes in the network or the environment, for example, if the density of IoT devices changes or if the IoT network expands.

%Based on (\ref{P4}) we can see that for parameter estimation we require a sufficient number of measurements of JDP between different pairs of locations across each band $m$. 
The required $\mathcal{S}_m$ estimates will be collected by the currently installed BSs and, if necessary, measurements collected by $\hat{B}$ temporary installed BSs that can assist with training.  
Installation of temporary BSs can be assisted with the help of UAVs that can carry and place the BSs to a designated location.
%\pp{ - PP: I brought this back, it is a nice observation.} 
During the operation of the network, installed BSs and temporary BSs will record transmissions that they receive, as well as their timestamps and the band that they were received on. Using the core network, BSs forward their recordings to a central processor, which then estimates the JDPs, $\tilde{\mathbb{E}}_r\{A_{b,m}(r)A_{v,m}(r)\}$. 
We assume that the central processor can identify all the successful transmissions that occurred from the captures. 
{Hence, the network can only have information about the successful transmissions, but not directly about the unsuccessful ones. In order to find the expected JDP, the infrastructure should know how many packets have been sent and how many repetitions have been made. This can be known by, for example, IoT devices adding packet and repetition numbering to their transmissions.}
%For the captures that were decoded by at least one BS, the transmissions can be identified based on the packet content, while for the minority of captures that have not been decoded by any BS but simply detected, this can be done based on the timestamps of their detection or based on IQ samples. 
%\pp{PP: The network can only have info on the successful transmissions, but not directly on the unsuccessful ones. In order to find the expcted value above, the infrastructure should know how many packets have been sent and how many repetitions have been made. This can be known by, e.g. putting a packet numbering plus a number of the repetitions. I think we need to put a comment on this in order to make the scheme implementable in practice.}

%tohere
The number of different $S_m$ estimates of JDP that can be collected for a given number of temporary and installed BSs is determined by their assignment to bands. For example, if there are $B_m$ BSs assigned to operate on a band $m$, then we can collect ${B_m}\choose{2}$ estimates of JDP on that band. 
However, if $S_m >$ ${B_m}\choose{2}$ estimates are needed for parameter estimation then it will be necessary to move BSs across bands to collect the necessary measurements. While the assignment of temporary BSs to bands can be changed without repercussions, the reassignment of installed BSs to bands will impact the PDP and therefore cannot be changed arbitrarily, since we assume that the temporary BSs do not contribute to the PDP. For example, we cannot assign all of the installed BSs to a single band during training as leaving the other bands to be unoccupied would severely harm the PDP.   

Let $\{1,\dots,B,\dots,\hat{B}+B\}$ be the set of locations of installed and temporary BSs combined. Then, let us use the assignment variable $\mathbf{X}\in \mathbb{Z}_2^{(\hat{B}+B) \times M}$ for the assignment of both temporary and installed BSs to bands, where $[\mathbf{X}]_{b,m}$ is equal to 1 if BS $b$ is assigned to band $m$ and 0 otherwise.
Furthermore, let the set of viable assignments during training be $ \mathcal{X}_{\text{train}}$, where $ \mathcal{X}_{\text{train}}=\{\mathbf{X}_i~|~i=1,\dots,|\mathcal{X}_{\text{train}}|\}$. 
The set $\mathcal{X}_{\text{train}}$ is determined by the constraints on the assignment of installed BSs. The constraints may be heuristic such as ensuring that a certain number of BSs remains on a particular band or ensuring to select assignments whose predicted TDP exceeds a certain threshold. The predicted TDP for a particular assignment $\mathbf{X}$ can be obtained by using estimates of ADP and JDP from earlier training episodes and evaluating the objective function in (\ref{P3}). 

Let us now define the training procedure. There will be several phases, during each of which a different assignment $\mathbf{X}_i$ will be applied. Since our objective is to complete the training as fast as possible, we need to select the least number of assignments $\mathbf{X}_i$ such that $S_m$ distinct JDP estimates are obtained on each band $m$. The set of assignments selected for training is $\mathcal{L}$. 

We observe that the problem of selecting $\mathcal{L}$ is a variant of the NP hard \emph{set cover problem}. In the set cover problem, given a collection of subsets $\mathcal{T}$ of a ground set $\mathcal{U}$, the goal is to cover $\mathcal{U}$ with the smallest number of subsets. An extension of the set cover problem is the \emph{partition set cover problem}, also an NP problem. In the partition set cover problem, $\mathcal{U}$ is divided into $M$ partitions, $\mathcal{U} = \{\mathcal{U}_1,\dots,\mathcal{U}_M\}$, and the goal is to cover $S_m$ elements from each $\mathcal{U}_m$ with the smallest number of subsets from $\mathcal{T}$. Let us demonstrate that the problem of selecting $\mathcal{L}$ is the partition set cover problem. Let $\mathcal{U}_m=\{\tilde{\mathbb{E}}_r\{A_{b,m}(r)A_{v,m}(r)\}~|~b=1,\dots,\hat{B}+B,~v=1,\dots,b \}$ be the set of distinct JDP estimates that can be collected on band $m$. 
Furthermore, let $\mathcal{T}_i$ be the set of JDP estimates that can be collected given the assignment $\mathbf{X}_i \in \mathcal{X}_{\text{train}}$, and $\mathcal{T}=\{\mathcal{T}_1,\dots,\mathcal{T}_{|\mathcal{X}_{\text{train}}|}\}$. 
We assume that $\cup_{i=1,\dots,|\mathcal{X}_{\text{train}}|}\mathcal{T}_i$ and $\cup_{m=1,\dots,M} \mathcal{U}_m$ are identical. If not, $\mathcal{U}$ can be revised to remove the JDPs $\mathcal{U} \setminus \cup_{i=1,\dots,|\mathcal{X}_{\text{train}}|}\mathcal{T}_i$. Selecting the $\mathcal{L}$ assignments is equivalent to selecting $|\mathcal{L}|$ sets $\mathcal{T}_i$ that will cover $S_m$ elements from each partition $\mathcal{U}_m$, therefore this is a partition set cover problem. For the sake of simplicity, in this paper, we will use a polynomial time greedy approach to solve the partition set cover problem developed in \cite{slavik1997improved}. {Better solutions can also } be pursued, such as the one in \cite{bera2014approximation}. We describe the greedy approach for planning of the training procedure in Algorithm \ref{alg:partition_set_cover}. 

\begin{algorithm}[t!]
\DontPrintSemicolon
%\SetAlgoLined
\KwIn{$\mathcal{U}$, $\mathcal{T}$, $S_1,\dots,S_M$}
\KwOut{$\mathcal{L}$}
$\mathcal{L}= \emptyset$

%$s_m \leftarrow S_m \forall m$

\While{any $s_m > 0$}{

Find $l\in(\{1,\dots,|\mathcal{X}_{\text{train}}|\} \setminus \mathcal{L})$ that maximizes $\min(\sum_m s_m, \sum_m |\mathcal{U}_m \cap \mathcal{T}_l|)$

$S_m \leftarrow S_m - |\mathcal{U}_m \cap (\cup_{l \in \mathcal{L}}\mathcal{T}_l)|~\forall m$

Add $l$ to $L$

$\mathcal{T}_i \leftarrow \mathcal{T}_i \setminus \mathcal{T}_l$ for $i=1,\dots,|\mathcal{X}_{\text{train}}|$

}

\caption{Greedy algorithm for the selection of $\mathcal{L}$ BS assignments $\mathbf{X}_i$ for training.}
\label{alg:partition_set_cover}
\end{algorithm}
The algorithm returns the set of assignments $\mathcal{L}$ that will result in $S_m$ estimates from $\mathcal{U}_m$ being collected for each $m$. 
Assignments are added to $\mathcal{L}$ with priority given to the assignments that can learn most new estimates in $\mathcal{U}$ given the currently selected assignments in $\mathcal{L}$.

\section{Measurement-based placement and assignment approach}
\label{sec:meas-approach}

In this section, we will discuss an alternative approach to MOD, in which ADPs, ${\mathbb{E}}_r\{A_{b,m}(r)\}~\forall b$, and JDPs ${\mathbb{E}}_r\{A_{b,m}(r)A_{v,m}(r)\}~\forall b,v,m$ are estimated directly using measurements from BSs in the field instead of predicting ADPs and JDPs via modeling. As in the MOD approach, the estimated ADPs and JDPs are plugged into (\ref{P3}) to solve for optimal placement and band assignment of BSs. We refer to this approach as MEAS for convenience. 
%The MEAS approach has a higher training complexity compared to the MOD approach since it needs to estimate a larger number of parameters, however, its advantage is that it has very loose assumptions on the model of the environment. There are no assumptions on the channel model or the spatial distribution of the IoT or incumbent nodes. Furthermore, in this approach there are no assumptions on the technology of the incumbent network such as their MAC protocols or their distribution across the multiplexing bands.
% This has now been explained earlier, so it commented out.
{\subsection{Training procedure}}
The training involves the direct estimation of the necessary ADPs, $\hat{\mathbb{E}}_r\{A_{b,m}(r)\}~\forall b,m$, and the necessary JDPs $\hat{\mathbb{E}}_r\{A_{b,m}(r)A_{v,m}(r)\}~\forall b,v,m$. Similar to the training phase of the MOD approach, during the operation of the network, installed BSs and temporary BSs will record transmissions that they receive, as well as their timestamps and the band that they were received on. BSs forward their recordings to a central processors, which then estimates the ADPs, $\hat{\mathbb{E}}_r\{A_{b,m}(r)\}~\forall b,m$, and JDPs $\hat{\mathbb{E}}_r\{A_{b,m}(r)A_{v,m}(r)\}~\forall b,v,m$. Evidently, the assignment of installed and temporary BSs will determine which ADPs and JDPs can be measured. 
Hence, the training may have to take place over several stages with a different BS band assignment applied in each. 

The planning of the training phases is similar problem to the one explained in Sec. \ref{sec:mod-training-procedure} for the MOD approach. The main difference is that $ \mathcal{U}$ includes all ADPs and all JDPs. 
The assumption remains that a certain set of assignments $ \mathcal{X}_{\text{train}}=\{\mathbf{X}_i~|~i=1,\dots,|\mathcal{X}_{\text{train}}|\}$ is possible during training and for each assignment $i$, a subset of measurements from $\mathcal{U}$, $\mathcal{T}_i$, can be collected. 
After, applying a similar analysis as in Sec. \ref{sec:mod-training-procedure}, the planning of the training phases can be posed as a set cover problem instead of a partition set cover problem. Likewise, the planning can be solved using the greedy approach described in Algorithm 2. %However, $S_m = |\mathcal{U}_m|~\forall m$, since all parameters from $\mathcal{U}$ must be estimated.  
Set cover solutions that are more optimal than a greedy approach can be applied and are abundant in literature. 
For simplicity, we implement the greedy solution since evaluating different set cover solution algorithms would be beyond the scope of this paper. 

{
\subsection{Comparison to model-based BS placement and band assignment}}

{The training process of the MEAS approach is similar to the training process of the MOD approach, however, there are some notable differences. First, the MEAS approach requires all locations in $\{1,\dots,B+C\}$ to be covered by either previously installed BSs or temporary BSs, since ADPs and JDPs are estimated through direct measurements rather than through modelling. 
On the other hand, the MOD approach only requires the ADP and JDP to be estimated for a subset of BS locations, and then uses modeling to predict ADP and JDP on the remainder of the locations. This means that fewer temporary BSs need to be installed at candidate locations for optimal infrastructure planning. For these reasons, the MEAS approach is more appropriate when $\Delta B = 0$, i.e.{,} when only performing frequency assignment, or when the number of candidate locations $C$ is small. Moreover, even if $\Delta B = 0$, if $B$ is large, the MEAS approach will suffer from an extended training time, since ${B}\choose{2}$ JDP values need to be estimated, whereas MOD approach generally requires less than ${B}\choose{2}$ JDP measurements, as we discussed in \ref{sec:mod-training-procedure} and \ref{sec:training-parameter-estimation}.
When $B$ is large or when $C$ is large and $\Delta B > 0$, MEAS approach is not practical, however it still serves as a useful benchmark in our simulation results, since it is an upper bound for the MOD approach in terms of the PDP.}

{The MOD approach has a lower training complexity and is more practical than the MEAS approach, however, the incumbent network and the IoT network are expected to match the system model explained in Sec. \ref{sec:iot-network-model} and additional channel modelling assumptions introduced in Sec. \ref{sec:model-based-approach}.
On the other hand, the MEAS approach only has to meet the modelling assumptions in Sec. \ref{sec:iot-network-model-a}.}
%Given that the MOD approach uses modeling to predict ADPs and JDPs, {an error in this prediction will cause a suboptimal deployment of BSs.} {The error in ADP and JDP prediction can arise due to model mismatch between the assumed model and the environment or due to an inaccurate model fit to the} environment. 

\section{Simulation results}
\label{sec:sim-results}

\begin{table}[t]
%\vspace{0pt}
\renewcommand{\arraystretch}{0.7}
\caption{Parameter values used in simulation}
\label{table:params}
\centering

\begin{tabular}{c||c||c||c}
\hline 
Parameter & Value & Parameter & Value\tabularnewline
\hline 
\hline 
Noise power & $-146$ dBm & $w_I $ & 200 kHz\tabularnewline
\hline 
$P_I$ & 14 dBm & $N_I$ & $3$\tabularnewline
\hline 
$P$ & 14 dBm & $R_I$ & $1$ \tabularnewline
\hline 
$R$ & 3 & Packet size incum. & $200$ B\tabularnewline
\hline 
$N$ & $3$  & $T_I$ & $200~B/w_I$ s \tabularnewline
\hline 
Packet size IoT & $20$ B & $B$ & 6 \tabularnewline
\hline 
$w$ & $600$ Hz& $\tau$  & $10$ dB\tabularnewline
\hline 
$T$ & $20~B/w$ s & $T_{\text{train}}$  & 10 min \tabularnewline
\hline 
$W$ & $200$ kHz & $M$  & $3$ \tabularnewline
\hline 
$\Lambda_{IoT}$ & 50 km$^{-2}$ & $\Lambda_{I}$  & 50 km$^{-2}$ \tabularnewline
\hline 

\end{tabular}

\end{table}

\subsection{Simulation environment}

We simulate a disk area of radius $10$ km with BSs, UNB devices and incumbent devices randomly distributed over the area. In each Monte Carlo (MC) simulation, the BS locations are sampled from a uniform distribution. Similarly, in each MC simulation, the UNB devices are sampled from a homogeneous Poisson Point process (HPPP) with a density $\Lambda_{IoT}$, { 
therefore, the expected number of devices across all Monte Carlo simulations is $\Lambda_{IoT}|\mathcal{A}|$, where $|\mathcal{A}|$ is the area size of the simulation area.}
The locations of incumbent devices are generated from a HPPP with density $\Lambda_{I}$.  Furthermore, the placement of current BSs and candidate locations is also randomized.
The distribution of incumbent devices across multiplexing bands is non-uniform. Let $\beta_j$ be the operating band of the incumbent device $j$.  The probability that an incumbent is assigned to operate in a multiplexing band $m$ during an MC iteration is a non-uniform discrete PMF ${Pr}(\beta_j = m)~\forall m$. The PMF ${Pr}(\beta_j = m)~\forall m$ distribution is randomly generated in each MC iteration.
When planning the training procedure of MOD and MEAS, we assume that $\mathcal{X}_{\text{train}}$ contains all assignments such that each band is covered by at least $\left\lfloor \frac{B}{M} \right\rfloor $ BSs.
Each MC iteration has a training stage for the MOD or MEAS approach. The training stage lasts $T_{\text{train}}$ minutes and is split equally across each training phase. After the training stage ends and the BS band assignment and BS placement are completed, the network runs for 1 h to collect the performance metrics. The results are averaged over 300 MC iterations.

{Unless otherwise stated, we use the simulation parameters
given in Table  \ref{table:params}, where the UNB network emulates the Sigfox network with US specifications \cite{sigfox_usa_2020}. The temporal traffic generation assumes that each device sends three unique packets per hour. For
interfering incumbents, we consider specifications similar to LoRa IoT devices \cite{lorawan} and assume they have a similar temporal traffic generation characteristics as the UNB network.
We note that the noise is not ignored, and is equal to the thermal noise power at room temperature across bandwidth of $w$  Hz.}

\subsection{BS band assignment results}
\label{sec:band-assignment-results}

% \begin{figure}[t]
%     \centering
%     \includegraphics[width=0.7\linewidth]{figures/detection_probability_vs_lambda_IoT.eps}
%     \captionof{figure}{The PDP of the assignment algorithms with respect to the density of the IoT devices in the network. }
%     \label{fig:pdp-density}    
% \end{figure}

\begin{figure*}[t]
    \centering
    \begin{minipage}{.49\textwidth}
    \centering
    %\vspace{-18pt}
    \includegraphics[width=0.85\linewidth]{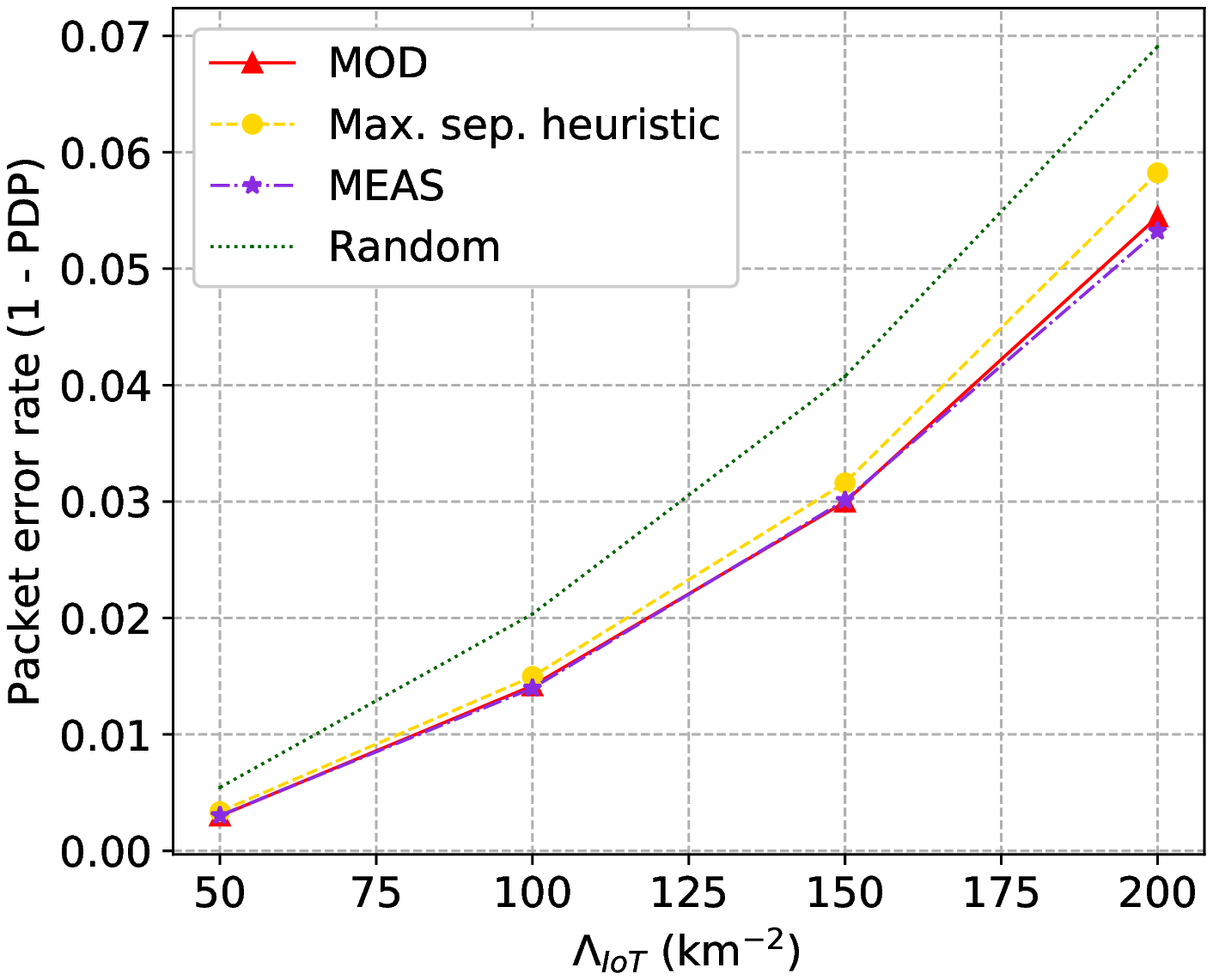}
    \captionof{figure}{The PDP of the assignment algorithms with respect to the density of the IoT devices in the network. }
    \label{fig:pdp-density}     
    \end{minipage}  
    \begin{minipage}{.01\textwidth}
    \end{minipage}
    \begin{minipage}{.49\textwidth}
    \centering
    \includegraphics[width=0.85\linewidth]{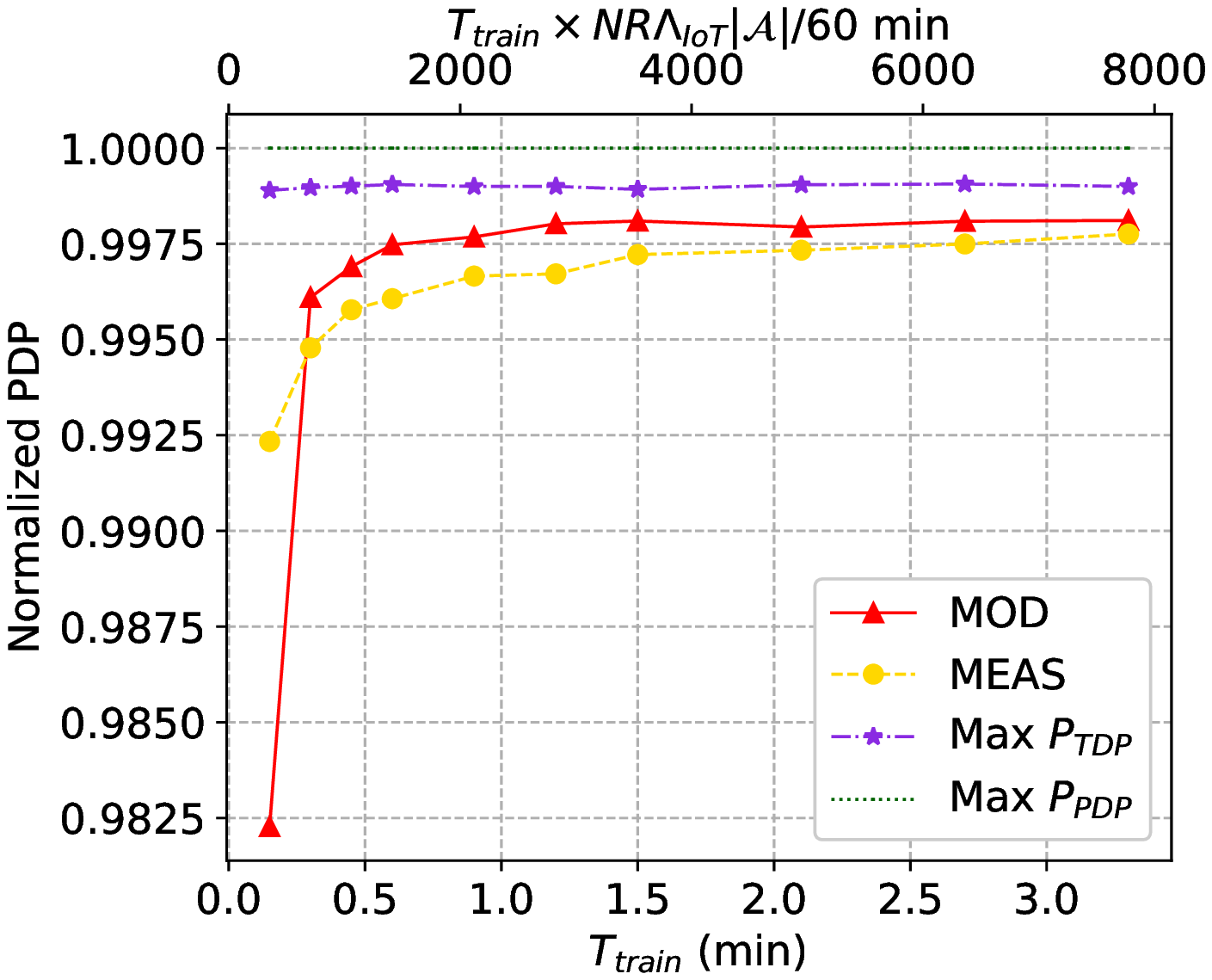}
    \captionof{figure}{The performance of the MOD and MEAS approach with respect to the length of the training time $T_{\text{train}}$. The PDP is normalized by the PDP of the theoretical $P_{\text{PDP}}$-maximizing assignment. }
    \label{fig:results-t-train}  
    \end{minipage}

\end{figure*}

\begin{figure*}[t]
    \centering
    \vspace{-10pt}
    \begin{minipage}{.49\textwidth}
    
    \centering
    \includegraphics[width=0.85\linewidth]{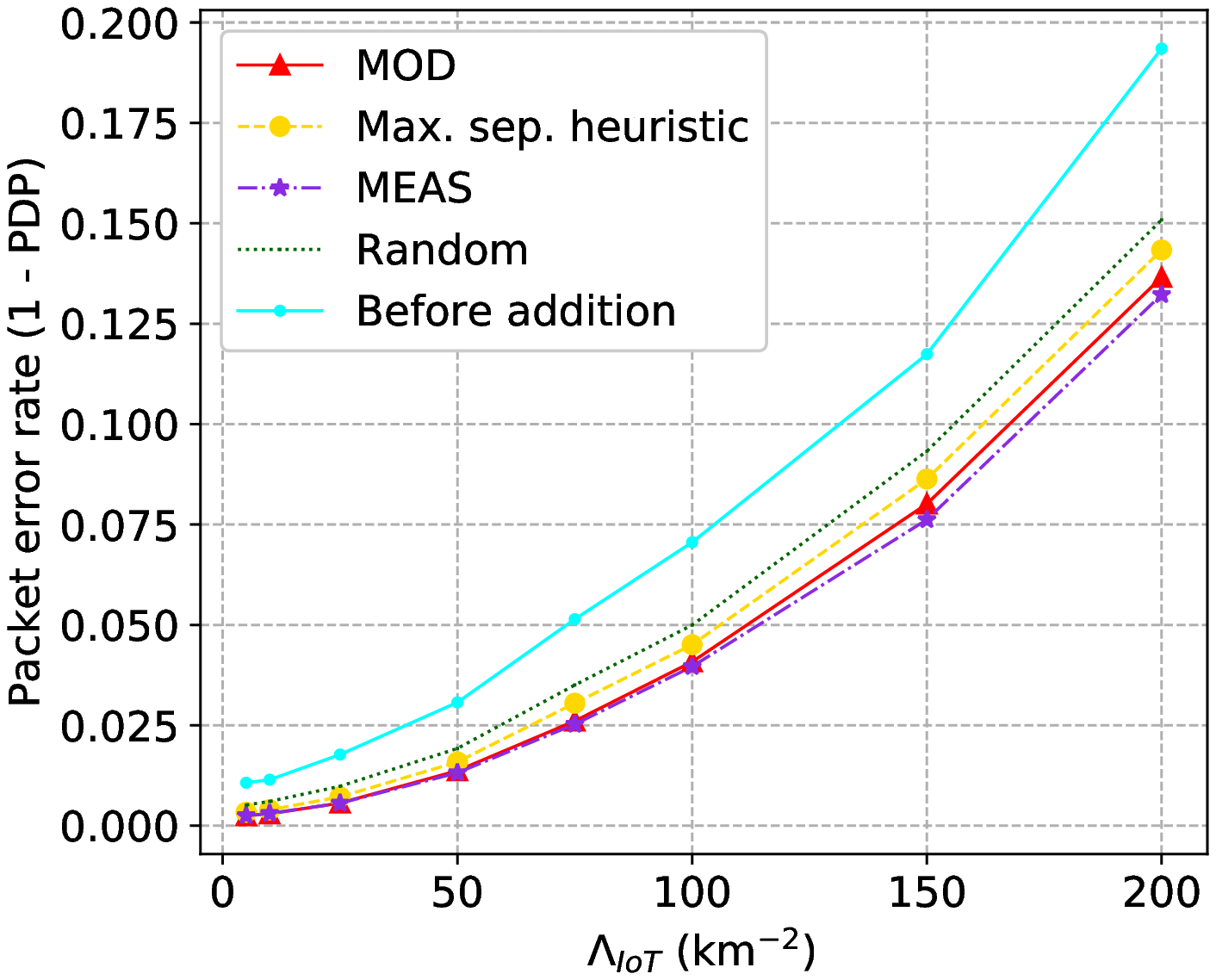}
    \caption{The PDP of the tested placement and assignment algorithms with respect to the density of the IoT devices in the network. }
    \label{fig:pdp-add}
    \end{minipage}
    \begin{minipage}{.01\textwidth}
    \end{minipage}
    \begin{minipage}{.49\textwidth}
    \centering
    \includegraphics[width=0.82\linewidth]{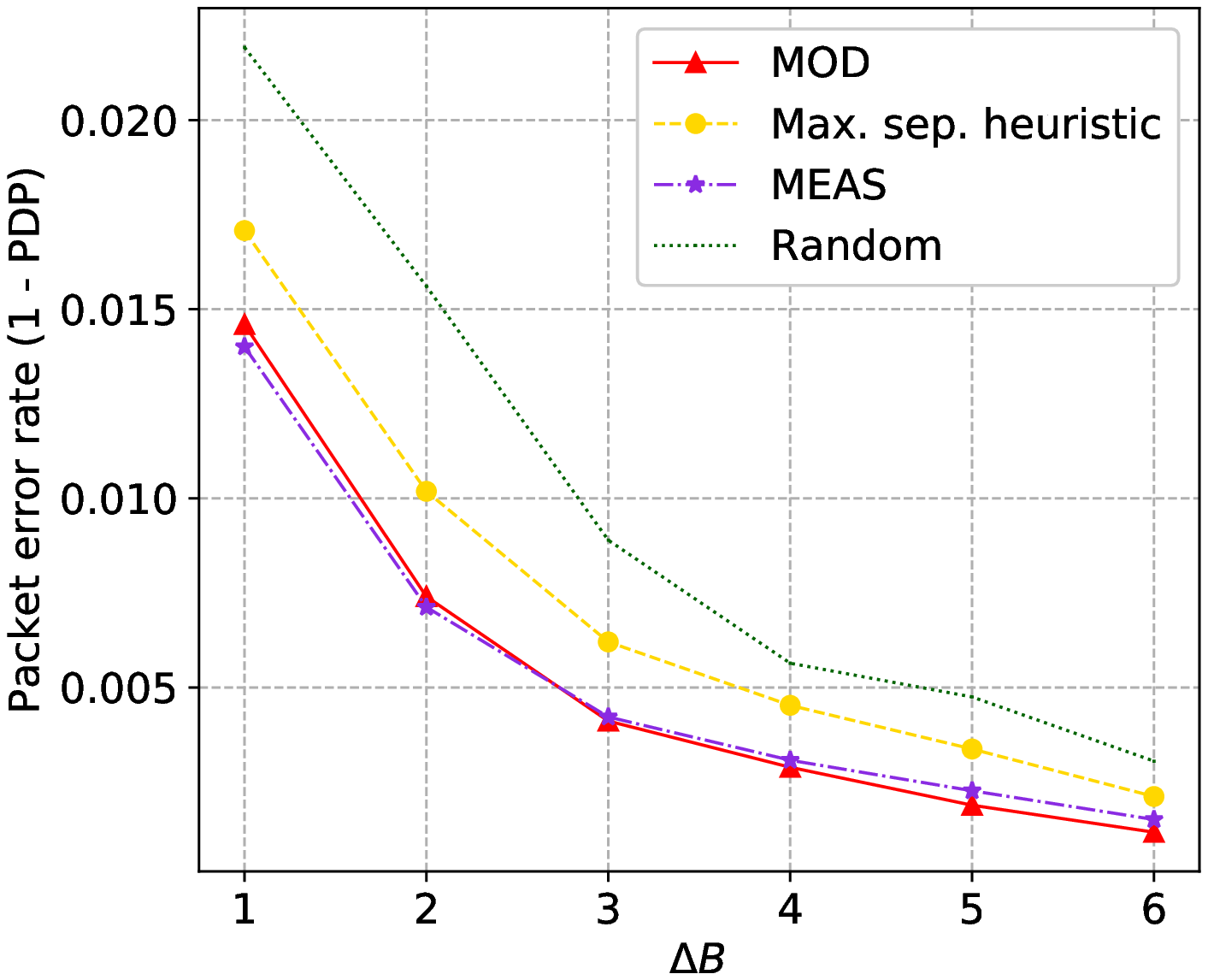}
    \caption{The improvement in PDP after $\Delta B$ new BSs are added to the network for different placement algorithms. }
    \label{fig:vary-delta-b}
    \end{minipage}
    \vspace{-10pt}
\end{figure*}

\label{sec:bs-assig-resul}
The following algorithms are evaluated only for BS assignment to bands in terms of the packet decoding probability (PDP): 
\begin{enumerate}
    \item Random assignment: The BSs are randomly assigned to bands with the constraint that each band is covered by at least $\left\lfloor \frac{B}{M} \right\rfloor $ BSs.
    {\item Maximum BS separation heuristic: The BSs are placed by solving the optimization problem
    \begin{equation*}
    \begin{aligned}
    \max_{\mathbf{X}} \quad & \sum_{m}\bigg(\sum_{b}
    \sum_{v<b}[\mathbf{X}]_{b,m}[\mathbf{X}]_{v,m}d_{b,v}\bigg) \\
     \textrm{s.t.} \quad & \text{(\ref{P1.2}), (\ref{P1.3}),  (\ref{P1.4})},
    \end{aligned}
    %\tag{P3}
    %\label{P3}
    \end{equation*}
    which ensures that BSs that are placed on the same band have maximum separation and that new BSs are added such that separation from the current BSs is maximized. This heuristics requires only knowledge of the coordinates of the current BSs as well as the  candidate locations for BSs. Maximizing the separation ensures that the BSs are not decoding packets from common transmissions, which may increase the network PDP. This heuristic may not always work, as UNB LPWANs require diversity in terms of BS reception when the decoding probability is low.}
    \item MOD assignment: No additional temporary BSs $\hat{B}$ are used in training. Furthermore, we set $S_m = 10~ \forall m$.
    \item MEAS assignment.
    \item Optimal assignment that maximizes the probability of decoding a transmission, $P_{\text{TDP}}$. The optimal assignment for maximum transmission decoding rate can be obtained in simulation. We rerun every MC realization, with its particular sequence of pseudo-random events, for all possible $M^B$ assignments, and select the assignment that gives the highest average packet or transmission decoding rate. 
    \item Optimal assignment that maximizes the probability of decoding a packet, $P_{\text{PDP}}$. The solutions are obtained through the same approach as for optimal $P_{\text{TDP}}$ assignment.
\end{enumerate} 

%\begin{figure}[t!]
%    \centering

%\end{figure}

First, we evaluate the performance of the algorithms with respect to the density of IoT devices. {Here, we set the number of BSs to $B=12$ and use high IoT device density, since with fewer BSs and low density, the band assignment problem becomes simpler and the gap between the heuristics and proposed solutions is smaller.} 
The results are shown in Fig. \ref{fig:pdp-density}. 
Typically, the PDP in real UNB networks such as Sigfox  is $>$95\%, therefore our simulation setting is realistic in that regard.
Random assignment performs significantly worse than either of our proposed approaches (MOD or MEAS). With assignment obtained using either of the proposed algorithms, the BS infrastructure can support around 25 more devices per km$^2$ in high density scenarios while maintaining the same PDP as random assignment, {and around 5 more devices compared to the maximum separation heuristic.} 
%However, the gap between $P_{\text{TDP}}$-maximizing assignment and $P_{\text{PDP}}$-maximizing assignment increases slightly with device density. 

We also analyze the dependence of the MOD and MEAS approach on the length of the training $T_{\text{train}}$. This is important because, when training is performed, the BSs need to move away from their optimal band assignment which may lower the current PDP and so the training phase should be as short as possible. The results are shown in Fig. \ref{fig:results-t-train}. We included a scaled axis, where $T_{\text{train}}$ is scaled by $NR\Lambda_\text{IoT}|\mathcal{A}| / 60$ min, which is equivalent to the expected number of IoT transmissions in $\mathcal{A}$ over time $T_{\text{train}}$. The higher the number of transmissions received by the BSs, the more accurate will the estimates of ADP and JDP be. From Fig. \ref{fig:results-t-train}, it can be observed that the MOD approach requires less training time to achieve its highest PDP compared to the MEAS approach. 
This is because in the training stage of the MOD approach fewer parameters need to be estimated than in the training stage of the MEAS approach and the number of training phases $|\mathcal{L}|$ for the MOD approach is smaller. 
Since $T_{\text{train}}$ is equally split between the $|\mathcal{L}|$ training phases, each training phase lasts longer in MOD approach and so the parameters estimated during each training phase contain less error. 
Moreover, the MOD and MEAS approach have a very similar performance to the theoretical best assignment that maximizes $P_{\text{TDP}}$. That indicates that the objective function in (\ref{P3}) is a tight lower bound to the objective function in (\ref{P2}). Furthermore, the theoretical best assignment that maximizes $P_{\text{TDP}}$ is very close in performance to the theoretical best assignment that maximizes $P_{\text{PDP}}$. This indicates that the objective function in (\ref{P2}) is a tight lower bound to the objective function in (P1).

\subsection{BS band assignment and placement results}

% \begin{figure}[t]
%     \centering
%     \includegraphics[width=0.7\linewidth]{figures/detection_probability_vs_lambda_IoT_add.eps}
%     \caption{The PDP of the tested placement and assignment algorithms with respect to the density of the IoT devices in the network. }
%     \label{fig:pdp-add}
% \end{figure}

The following algorithms are evaluated for both BS placement and BS assignment to bands in terms of the PDP: 
\begin{enumerate}
    \item Random placement with optimized band assignment: The $\Delta B$ BSs are randomly placed within the $C$ candidate locations.  With the placement of $\Delta B$ BSs decided, the band assignment of $B + \Delta B$ is done using the MEAS approach.
    \item Maximum BS separation heuristic: As described in Sec. \ref{sec:band-assignment-results}
    \item MOD assignment: No additional temporary BSs $\hat{B}$ are used in training and $S_m = 10~ \forall m$.
    \item MEAS assignment: This approach is not practical for placement of BSs since it involves placing a temporary BS at each of the $C$ candidate locations, but we include it as a benchmark for the MOD approach.
\end{enumerate} 
The theoretical optimal assignment for maximum packet decoding rate or maximum transmission decoding rate cannot be obtained numerically in simulation as with BS assignment in Sec. \ref{sec:bs-assig-resul} since searching for the optimal assignment and placement exhaustively is intractable due to the number of possible solutions for BS placement and assignment.

The number of candidate locations is $C = 30$ and their locations are uniformly and randomly distributed in each MC iteration. The number of new BSs being installed is {$\Delta B = 1$}. We evaluate the performance with respect to the density of IoT devices and the results are shown in Fig. \ref{fig:pdp-add}. Before addition of the BSs the frequency assignment of BSs is obtained using the MEAS approach. 
Putting additional BSs significantly improves the PDP, however, with optimized placement of the new BSs, additional improvements can be achieved. Using the MOD approach additional 20 devices per km$^{2}$ can be supported while maintaining a similar PDP as with random placement, {and around 5 more compared to the maximum separation heuristic.} 
Furthermore, the MOD approach is similar in PDP to the MEAS approach which is a good demonstration of the practical usefulness of the MOD approach. As a reminder, the MEAS approach results are obtained by placing temporary BSs at each of the $C$ candidate locations to estimate ADPs and JDPs while the MOD approach predicts the ADPs and JDPs at candidate locations using only the measurements from the currently installed $B$ BSs. Despite of this, the PDP obtained when using MOD placement and assignment is the same as the PDP when using the MEAS placement and assignment. In Fig. \ref{fig:vary-delta-b} we increase the number of candidate positions to $C=50$ and vary $\Delta B$ from 1 to 6.  As $\Delta B$ increases, the difference between random and optimized placement becomes smaller. Therefore, optimized placement becomes more important when fewer new BSs can be afforded to be installed in the IoT network. {Furthermore, we can see that the proposed solutions can achieve the same PDP as the random heuristic with two fewer BSs, and one fewer BS compared to the maximum BS separation heuristic.}
Additionally, we observe that MOD approach outperforms the MEAS approach for some $\Delta B$, which can be attributed to the MEAS approach not having sufficient training time to estimate ADP and JDP accurately enough.

% Finally, we show how the assignment affects devices depending on how far away they are from base stations. We measure the decoding probability as dependent on the mean distance of a user to all basestations in the network. These results are shown in Fig. \ref{fig:distance}. Naturally, for users that are farther away from all basestations the decoding probability is lower. Performing optimal assignment rather than just applying random assignment is more beneficial for users that are farther away from all basestations than for those that are close. This makes sense since the users that are close to all basestations already have enough diversity in terms of the number of basestations that detect their packets, therefore there is not much room for improvement.

% \begin{figure}[h!]
% 	\centering
% 	\includegraphics[width=0.9\linewidth]{figures/prob_mean_distance.pdf}
% 	\caption{The performance of the proposed algorithm depending on the mean distance of a user to all basestations in the network.}
% 	\label{fig:distance}

% \end{figure}

\section{Conclusions}
\label{sec:conclusions}

In this work, we have addressed BS infrastructure management and expansion in multi-band UNB IoT networks. Namely, we devise two algorithms for maximization of packet delivery through optimal placement of BSs and optimal assignment of BSs to multiplexing bands. We derive a model-based and a measurement-based algorithm. The model-based approach has a simpler training stage than the measurement-based algorithm, however, it is based on certain assumptions that the IoT network and the environment should match. In the development of the model-based algorithm, we derive models of average decoding probability and joint decoding probability using stochastic geometry which we verify through simulation. Furthermore, our assignment and placement algorithms offer significant improvement in packet decoding probability over baseline approaches and in certain cases closely match the theoretical best performance. Future work will focus on optimizing the access policy of IoT devices in a multi-band UNB network and will also consider collaborative decoding of transmission by several BSs through fusion of their received signals.
\appendix
\subsection{Proof of Proposition \ref{prop:suboptimal}}
\label{appendix:suboptimal}
Since we assume $\Pr(\beta(r)=m)$ is constant across $m$, we can rewrite the transmission decoding probability as
$P_{\text{TDP}}(\mathbf{X})=\sum_m\Pr\left(\bigcup_{b}\left[\mathbf{X}\right]_{b,m}\text{SINR}_{b,m}(r)>\tau\right)$. 
Using the exclusion-inclusion principle we can express this function as:
\begin{multline}
    P_{\text{TDP}}(\mathbf{X})=\sum_{m}\Bigg(\sum_{b}\Pr\left(\left[\mathbf{X}\right]_{b,m}\text{SINR}_{b,m}(r)\geq\tau\right)-\\
    \sum_{v<b}\Pr\left(\left[\mathbf{X}\right]_{b,m}\text{SINR}_{b,m}(r)\geq\tau,\left[\mathbf{X}\right]_{v,m}\text{SINR}_{v,m}(r)\geq\tau\right)+\\
    ...+(-1)^{M-1}\Pr\left(\bigcap_{b=1}^{B}\left[\mathbf{X}\right]_{b,m}\text{SINR}_{b,m}(r)\geq\tau\right)\Bigg)
    \label{eq:inlude-exclude}
\end{multline}
%We can take the entries of $\mathbf{X}$ out of the probability functions to get the expression:
%\begin{multline}
%    P(\mathbf{X})=\sum_{m}\Bigg(\sum_{b}X_{b,m}\mathbb{P}\left(\gamma_{b,m}(n)\geq\tau\right)-\\
%    \sum_{b<k}X_{b,m}X_{k,m}\mathbb{P}\left(\gamma_{b,m}(n)\geq\tau,\gamma_{k,m}(n)\geq\tau\right)+\\
%    ...+(-1)^{M-1}\sum_{b<...<k}\mathbb{P}\left(\bigcap_{b=1}^{K}X_{b,m}\gamma_{b,m}(n)\geq\tau\right)\Bigg)
%    \label{eq:expansion}
%\end{multline}
The above sum can be truncated up to the second-order terms
\begin{multline}
P_{\text{TDP}}(\mathbf{X})\approx \sum_{m}\Bigg(\sum_{b}\left[\mathbf{X}\right]_{b,m}\Pr\left(\text{SINR}_{b,m}(r)\geq\tau\right)-\\
\sum_{b<v}\left[\mathbf{X}\right]_{b,m}\left[\mathbf{X}\right]_{v,m}\Pr\left(\text{SINR}_{b,m}(r)\geq\tau,\text{SINR}_{v,m}(r)\geq\tau\right) \Bigg)
\label{eq:tdp-approx}
\end{multline}
The approximation above is a concave function and is also a lower bound on the objective function in (\ref{P2}). The latter follows from Bonferroni inequalities, since we are taking a second order approximation of the expression in (\ref{eq:inlude-exclude}) \cite{bonferroni1936teoria}. 
In (\ref{eq:tdp-approx}), $Pr\left(\text{SINR}_{b,m}(r)\geq\tau\right) \equiv \mathbb{E}_r\{A_{b,m}(r)\}$ and $Pr\left(\text{SINR}_{b,m}(r)\geq\tau,\text{SINR}_{v,m}(r)\geq\tau\right) \equiv \mathbb{E}_r\{A_{b,m}(r)A_{v,m}(r)\}$. 
{The objective function can be proven to be concave by calculating the Hessian of the expression over $\text{vec}\left({\mathbf{X}}\right)$. The Hessian matrix will be a symmetric block diagonal matrix $\text{diag}\left(\mathbf{A}_{1},\dots, \mathbf{A}_{M}\right)$}
{, where each matrix $\mathbf{A}_{m}$ is a covariance matrix $\mathbb{E}_r\{\mathbf{a}(r)\mathbf{a}(r)^T\}$, of a random vector $\mathbf{a}(r)=\left[A_{1,1}(r), A_{1,2}(r), A_{B+C,M}(r)\right]^T$. Since covariance matrices are known to be positive semi-definite, including their sampled versions, the Hessian matrix is negative semi-definite and therefore the expression in (\ref{eq:tdp-approx}) is concave. 
}

\subsection{Proof of Theorem \ref{theorem:ccdf-sinr}}
\label{proof:ccdf-sinr}
The  complementary cumulative distribution (CCDF) of $\text{SINR}_{i,b,m}$ is 
\begin{multline*}
\Pr(\text{SINR}_{i,b,m}\geq\tau\mid p_{i,b}) \\
\stackrel{(a)}{=}
\Pr\left(h_{i,b}\geq s_{i,b}\left(\sum_{j \in \Phi}f_jp_{j,b}^{-\alpha} + \sum_{j' \in {\Phi}_{I,m}}
\hat{P}_I f_{j'}p_{j',b}^{-\alpha}\right)\right) \\
\stackrel{(b)}{=} \exp\left({s_{i,b}\left(\sum_{j \in \Phi}f_jp_{j,b}^{-\alpha}+\sum_{j' \in {\Phi}_{I,m}}\hat{P}_If_{j'}p_{j',b}^{-\alpha} \right)}\right)
\end{multline*}
where, to obtain $(a)$, we have assumed that the $\text{SINR}_{i,b,m}$ is dominated by interference over the noise (we set $\hat{P}_N \rightarrow 0$) and set $s_{i,b}=\tau p_{i,b}^{\alpha}$; $(b)$ follows from $h_i$ being modelled as  Exp$(1)$. Using a similar derivation as in \cite[Eq.~39]{hattab2018spectrum}, we can arrive at the expression 
\begin{equation*}
\Pr(\text{SINR}_{i,b,m}\geq\tau\mid p_{i,b}) 
= \exp\left(-\epsilon_m\tau^{\frac{2}{\alpha}}p_{i,b}^{2}\right),
\end{equation*}
where $\epsilon_m = \pi\left(\lambda+\hat{P}_{I}^{\frac{2}{\alpha}}{\lambda}_{I,m}\right)\frac{2\pi/\alpha}{\sin(2\pi/\alpha)}$. The above expression provides us with a complimentary CDF (CCDF) of the $\text{SINR}_{i,b,m}$ for a device $i$ with a given separation between the receiver $b$ and IoT device $j$, $p_{i,b}$. Note that in the derivation of the expression, we assumed that interference comes from IoT and incumbent devices that extend beyond $\mathcal{A}$ such that $\max_{j \in {\Phi}_{I,m}} p_{j,b} \rightarrow \infty$ and $\max_{j' \in {\Phi}_{m}}p_{j',b} \rightarrow \infty$. This approximation is reasonable if the BS $b$ is very far away from furthest away devices that are causing interference.

\subsection{Proof of Theorem \ref{theorem:joint-pdp}}
\label{proof:joint-pdp}
The distance of the device $i$ to BS $b$ is $r^2_{i,b}=r^2_{i}+\frac{d_{b,v}^2}{4}+p_{i}d_{b,v}\cos(\theta_i)$ and the distance to BS $v$ is $r^2_{i,v}=r^2_{i}+\frac{d_{b,v}^2}{4}-p_{i}d_{b,v}\cos(\theta_i)$. We define $s_{i,b}=\tau p_{i,b}^{\alpha}$ and $s_{i,v}=\tau p_{i,v}^{\alpha}$. The conditional joint decoding probability is then 
\begin{multline*}
\Pr\left(\text{SINR}_{i,b,m}\geq\tau,\text{SINR}_{i,v,m}\geq\tau\mid i\right) \\
%=
%Pr\Bigg(h_{i,b}\geq\tau p_{i,b}^{\alpha}\left(\sum_{j \in \Phi_{IoT}}f_{j,b}p_{j,b}^{-\alpha} + \sum_{j' \in \Phi_{I,m}}P_If_{j',b}p_{j',b}^{-\alpha}\right),\\
%h_{i,v}\geq\tau r_{i,v}^{\alpha}\left(\sum_{j \in \Phi_{IoT}}f_{j,v}p_{j,v}^{-\alpha} + \sum_{j' \in \Phi_{I,m}}P_If_{j',v}p_{j',v}^{-\alpha}\right) \Bigg) \\
\stackrel{(a)}{=}\exp\Bigg({s_{i,b}\left(\sum_{j \in \Phi_{IoT}}f_j p_{j,b}^{-\alpha}+\sum_{j' \in \Phi_{I,m}}\hat{P}_I f_{j'}p_{j',b}^{-\alpha} \right)} + \\ s_{i,v}\left(\sum_{j \in \Phi_{IoT}}f_{j,v}p_{j,v}^{-\alpha} + \sum_{j' \in \Phi_{I,m}}\hat{P}_If_{j',v}p_{j',v}^{-\alpha}\right) \Bigg) \\
%\stackrel{(b)}{=}\prod_{j\in\Phi_{IoT}}\left(\frac{1}{1+\tau p_{i,b}^{\alpha}p_{j,b}^{-\alpha}}\right)\left(\frac{1}{1+\tau r_{i,v}^{\alpha}p_{j,v}^{-\alpha}}\right) \times \\
%\prod_{j'\in\Phi_{IoT}}\left(\frac{1}{1+\tau P_{I}p_{i,b}^{\alpha}p_{j',b}^{-\alpha}}\right)\left(\frac{1}{1+\tau P_{I}r_{i,v}^{\alpha}p_{j',v}^{-\alpha}}\right) \\
\stackrel{(b)}{=} \exp \Bigg(\lambda \int_{\phi_j}\int_{p_j}\left(p_j-\frac{p_j}{\left(1+s_{i,b}p_{j,b}^{-\alpha}\right)\left(1+s_{i,v}p_{j,v}^{-\alpha}\right)}\right)  \\
+\lambda_{I,m}\int_{\phi_{j'}}\int_{p_{j'}}\left(p_{j'}-\frac{p_{j'}}{\left(1+s_{i,b}p_{j',b}^{-\alpha}\right)\left(1+s_{i,v}p_{j',v}^{-\alpha}\right)}\right)\Bigg)
\end{multline*}
where (a) follows from the assumption that the fading  RVs at BS $b$ and $v$ are independent; (b) follows from the expression for Laplace–Stieltjes transform of the Rayleigh RV and the PGFL of a HPPP. Evaluating the integral to a closed form expression is not possible to the best of our knowledge, therefore we adopt an approach similar to \cite{ganti2009spatial} to obtain the upper bound on the conditional joint outage probability. 

First, we will derive the expression for conditional decoding probability over the decoding probability of one of the BSs $\frac{\Pr(\text{SINR}{}_{i,v,m}\geq\tau\mid\text{SINR}_{i,b,m}\geq\tau)}{\Pr(\text{SINR}_{i,v,m}\geq\tau)}=\frac{\Pr(\text{SINR}_{i,b,m}\geq\tau,\text{SINR}_{i,v,m}\geq\tau)}{\Pr(\text{SINR}_{i,b,m}\geq\tau)\Pr(\text{SINR}_{i,v,m}\geq\tau)}$. Following the same derivation as in the expression above we can arrive at the product of two integrals:
\begin{multline*}
\frac{\Pr(\text{SINR}{}_{i,v,m}\geq\tau\mid\text{SINR}_{i,b,m}\geq\tau)}{\Pr(\text{SINR}_{i,v,m}\geq\tau)} = \\ \exp\left(\lambda\int_{\phi_j}\int_{p_j}\left(\frac{s_{i,b}p_{j,b}^{-\alpha}}{1+s_{i,b}p_{j,b}^{-\alpha}}\right)\left(\frac{s_{i,v}p_{j,v}^{-\alpha}}{1+s_{i,v}p_{j,v}^{-\alpha}}\right)\right) \times \\
\exp\left(\lambda_{I,m}\int_{\phi_j'}\int_{p_j'}\left(\frac{s_{i,b}p_{j',b}^{-\alpha}}{1+s_{i,b}p_{j',b}^{-\alpha}}\right)\left(\frac{s_{i,v}p_{j',v}^{-\alpha}}{1+s_{i,v}p_{j',v}^{-\alpha}}\right)\right)
\end{multline*}
If we let $d_{b,v} \rightarrow 0$, we can obtain an upper bound on the expression above and then evaluate the resulting integral to obtain:
\begin{equation*}
    \frac{\Pr(\text{SINR}{}_{i,v,m}\geq\tau\mid\text{SINR}_{i,b,m}\geq\tau)}{\Pr(\text{SINR}_{i,v,m}\geq\tau)} \leq 
    e^{\left(\left(-\frac{2}{\alpha}+1\right)\epsilon_m\tau^{\frac{2}{\alpha}}r^{2}\right)}.
\end{equation*}
Using the above expression we can then also derive the upper bound to the joint decoding probability:
\begin{multline*}
\Pr\left(\text{SINR}_{i,b,m}\geq\tau,\text{SINR}_{i,v,m}\geq\tau\right) = \Pr(\text{SINR}_{i,b,m}\geq\tau) \\
\times \Pr(\text{SINR}_{i,v,m}\geq\tau)  \frac{\Pr(\text{SINR}{}_{i,v,m}\geq\tau\mid\text{SINR}_{i,b,m}\geq\tau)}{\Pr(\text{SINR}_{i,v,m}\geq\tau)} \leq \\
\exp\left(\left(-\frac{2}{\alpha}-1\right)\epsilon_m\tau^{\frac{2}{\alpha}}r^{2}\right) \exp\left(-\frac{1}{2}\epsilon_m\tau^{\frac{2}{\alpha}}d_{b,v}^{2}\right).
\end{multline*}

\bibliography{references}
\bibliographystyle{ieeetr}
\vspace{-30pt}
\begin{IEEEbiography}[{\includegraphics[width=1in,height=1.25in,clip,keepaspectratio]{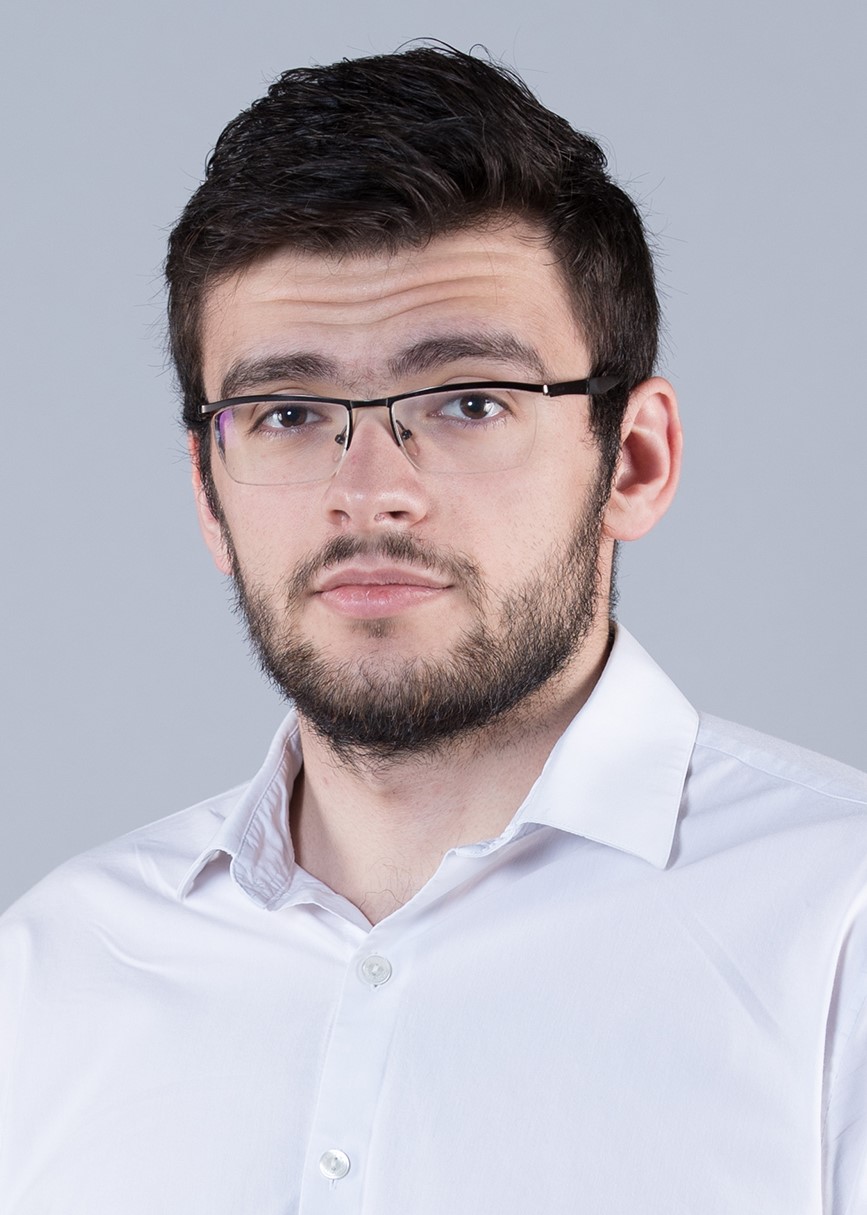}}]{Enes Krijestorac}
	received a B.S. degree in Electrical Engineering from New York University, Abu Dhabi in 2018, graduating \textit{summa cum laude}. He received his M.S. degree from University of California, Los Angeles, in 2021, also in Electrical Engineering. He is currently pursuing a Ph.D. degree at the University of California, Los Angeles, US. His research interests include UAV assisted wireless communication, wireless channel prediction and sensing, distributed computing over wireless channels and machine learning. He was a recipient of the Dean's Scholar Award and the PhD Preliminary Exam Fellowship at the UCLA Electrical and Computer Engineering Department. 
\end{IEEEbiography}

\begin{IEEEbiography}[{\includegraphics[width=1in,height=1.25in,clip,keepaspectratio]{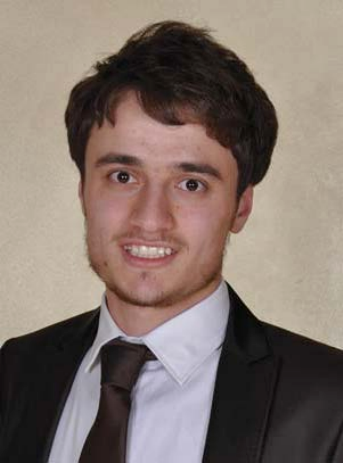}}]{Ghaith Hattab}
	received his PhD degree from the University of California, Los Angeles (UCLA). He received his M.Sc degree from Queen's University, Kingston ON, Canada, and his B.Sc degree from American University of Sharjah (AUS), Sharjah, United Arab Emirates (UAE), all in Electrical Engineering. He has held research internships at Nokia Bell Labs, IL, USA, and Qualcomm Research, CA, USA. He is currently with Apple Inc., USA. His research interests include system-level analysis, optimization, and signal processing for wireless communications, with emphasis on channel state feedback, spectrum sharing, and coexistence of wireless systems. He was the recipient of the prestigious Presidential Cup awarded by His Highness Sheikh Dr. Sultan Al Qassimi, the IEEE Kingston Section M.Sc Research Excellence Award, and the departmental fellowship at UCLA.
\end{IEEEbiography}

\begin{IEEEbiography}[{\includegraphics[width=1in,height=1.25in,clip,keepaspectratio]{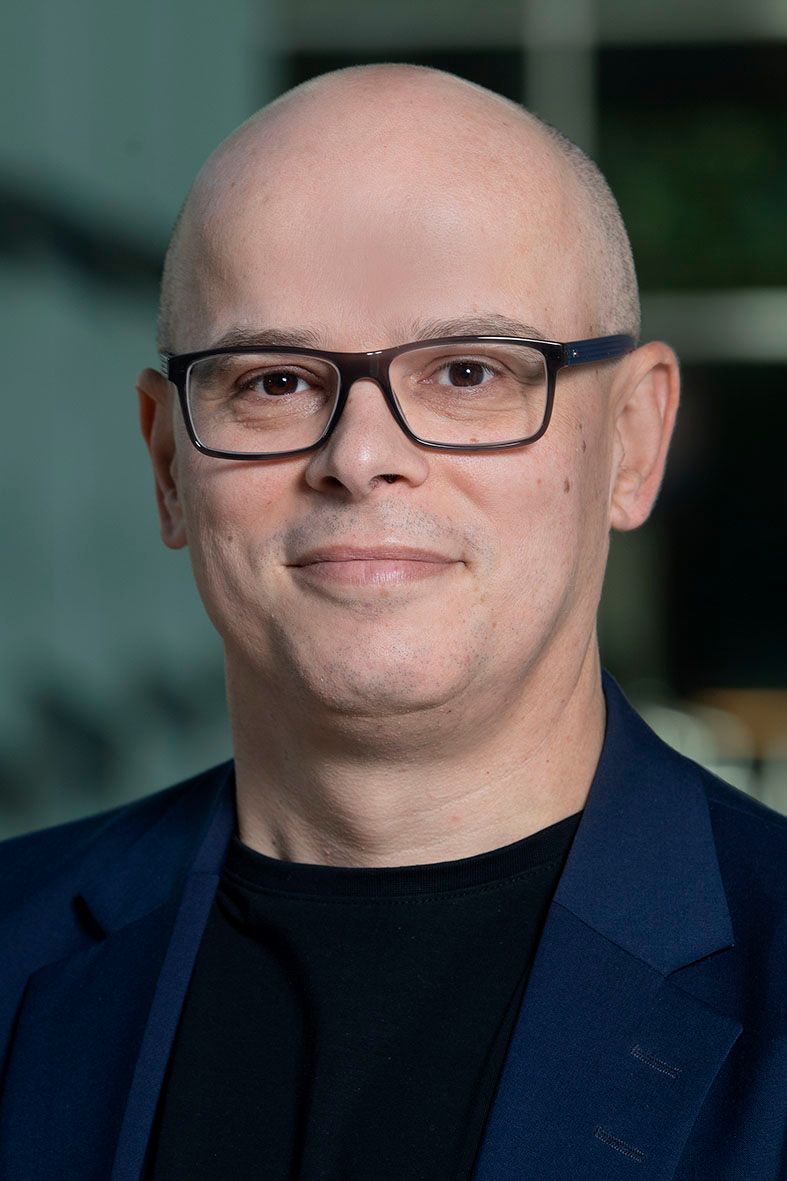}}]{Petar Popovski} (Fellow, 2016) is a Professor at Aalborg University, where he heads the section on Connectivity and a Visiting Excellence Chair at the University of Bremen. He received his Dipl.-Ing and M. Sc. degrees in communication engineering from the University of Sts. Cyril and Methodius in Skopje and the Ph.D. degree from Aalborg University in 2005. He received an ERC Consolidator Grant (2015), the Danish Elite Researcher award (2016), IEEE Fred W. Ellersick prize (2016), IEEE Stephen O. Rice prize (2018), Technical Achievement Award from the IEEE Technical Committee on Smart Grid Communications (2019), the Danish Telecommunication Prize (2020) and Villum Investigator Grant (2021). He was a Member at Large at the Board of Governors in IEEE Communication Society 2019-2021. He is currently an Editor-in-Chief of IEEEE JOURNAL ON SELECTED AREAS IN COMMUNICATIONS. He also serves as a Vice-Chair of the IEEE Communication Theory Technical Committee and the Steering Committee of IEEE TRANSACTIONS ON GREEN COMMUNICATIONS AND NETWORKING. Prof. Popovski was the General Chair for IEEE SmartGridComm 2018 and IEEE Communication Theory Workshop 2019. His research interests are in the area of wireless communication and communication theory. He authored the book ``Wireless Connectivity: An Intuitive and Fundamental Guide'', published by Wiley in 2020.
\end{IEEEbiography}

\begin{IEEEbiography}[{\includegraphics[width=1in,height=1.25in,clip,keepaspectratio]{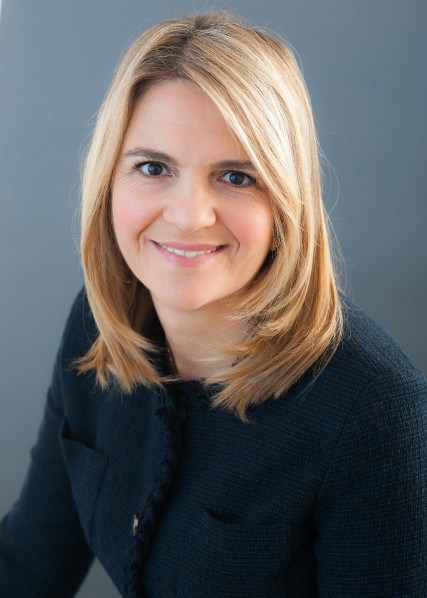}}]{Danijela Cabric}
	is Professor in Electrical and Computer Engineering at University of California, Los Angeles. She earned MS degree in Electrical Engineering in 2001, UCLA and Ph.D. in Electrical Engineering in 2007, UC Berkeley, Dr. Cabric received the Samueli Fellowship in 2008, the Okawa Foundation Research Grant in 2009, Hellman Fellowship in 2012 and the
	National Science Foundation Faculty Early Career Development (CAREER) Award in 2012 and the Qualcomm Faculty award in 2020 and 2021. She served as an Associate Editor of IEEE Transactions of Cognitive Communications and Networking, IEEE Transactions of Wireless Communications, IEEE Transactions on Mobile Computing and IEEE Signal Processing Magazine, and IEEE ComSoc Distinguished Lecturer. Her research interests are millimeter-wave communications, distributed communications and sensing for Internet of Things, and machine learning for wireless networks co-existence and security. She is an IEEE Fellow.
\end{IEEEbiography}

\end{document}